\documentclass{aa} 

\usepackage{graphicx}
\usepackage{txfonts}
\usepackage{natbib}
\bibpunct{(}{)}{;}{a}{}{,}

\newcommand{\be}{\begin{equation}}
\newcommand{\ee}{\end{equation}}
\newcommand{\angstrom}{\mbox{\normalfont\AA}}

\def\apjl{ApJL}
\def\apjs{ApJS}

\newcommand{\Mpc}{$h^{-1}$\thinspace Mpc}

\newcommand{\vmh}{h^{-1}\mathrm{Mpc} }

\DeclareMathAlphabet{\pazocal}{OMS}{zplm}{m}{n}

\begin{document}  

\title{
Multiscale cosmic web detachments, connectivity, and preprocessing \\
in the supercluster SCl~A2142 cocoon 
} 

\author {Maret~Einasto\inst{1} 
\and Boris Deshev\inst{2}
\and Peeter~Tenjes\inst{1} 
\and Pekka~Hein\"am\"aki\inst{3}
\and Elmo~Tempel\inst{1} 
\and Lauri Juhan~Liivam\"agi\inst{1} 
\and Jaan~Einasto\inst{1,4,5}
\and Heidi~Lietzen\inst{1}
\and Taavi~Tuvikene\inst{1}
\and Gayoung~Chon\inst{6}
}
\institute{Tartu Observatory, University of Tartu, Observatooriumi 1, 61602 T\~oravere, Estonia
\and
Astronomical Institute, Czech Academy of Sciences, 
Bocní II 1401, CZ-14131 Prague, Czech Republic
\and 
Tuorla Observatory, Department of Physics and Astronomy, University of Turku, 20014
Turku, Finland
\and
Estonian Academy of Sciences, Kohtu 6, 10130 Tallinn, Estonia
\and
ICRANet, Piazza della Repubblica 10, 65122 Pescara, Italy
%\and
%Max-Planck-Institut f\"ur extraterrestrische Physik, 85748 Garching, Germany
\and
Universit\"ats-Sternwarte M\"unchen, Fakult\"at f\"ur Physik, 
Ludwig-Maximilian-Universit\"at M\"unchen, Scheinerstr. 1, 81679 M\"unchen, Germany
}

\authorrunning{Einasto, M. et al. }

\offprints{Einasto, M.}

\date{ Received   / Accepted   }

\titlerunning{SCl~A2142env}

\abstract
{
Superclusters of galaxies and their surrounding low-density regions 
(cocoons) represent
dynamically evolving environments in which galaxies and their systems
form and evolve. While
evolutionary processes of galaxies in dense environments are extensively studied at present, 
galaxy evolution in low-density regions 
has received less attention.
}
{We study the properties, connectivity, and galaxy content 
of groups and filaments in the  A2142 supercluster 
(SCl~A2142) cocoon 
to understand the evolution of the supercluster with
its surrounding structures and the  galaxies
within them. 
}
{
We calculated the luminosity-density field 
of SDSS galaxies and traced the SCl~A2142 cocoon
boundaries by the lowest luminosity-density regions that separate SCl~A2142
from other superclusters.
We determined galaxy filaments and groups in the cocoon
and analysed the connectivity of groups, the high density core (HDC) 
of the supercluster, and the whole of the supercluster.
We compared the distribution and properties of 
galaxies with different star-formation properties
in the supercluster and in the cocoon. 
}
{The supercluster A2142 and  the long filament that is connected to it forms
the longest straight structure in the Universe detected so far, with a 
length of approximately $75$~\Mpc.
The connectivity of the cluster A2142 and the whole supercluster is $\pazocal{C} = 6 - 7$;
poor groups exhibit $\pazocal{C} = 1 - 2$.
Long filaments around the supercluster's main body are detached from it 
at the turnaround region.
Among various local and global environmental trends
with regard to the properties of galaxies and groups, we find that
galaxies with very old stellar populations 
lie in systems across a wide range of richness from the richest
cluster to poorest groups and single galaxies.
They lie even at local densities 
as low as $D1 < 1$ in the cocoon and up to $D1 > 800$ in the supercluster. 
Recently quenched
 galaxies lie in the cocoon mainly 
in one region and
their properties are different in the cocoon and in the supercluster. 
The star-formation properties of single galaxies 
are similar across all environments.
}
{
The collapsing main body of SCl~A2142 with the detached long filaments near it
are evidence of an important epoch in the supercluster 
evolution.
There is a need for further studies to explore possible reasons behind the similarities between galaxies with
very old stellar populations in extremely different environments,
as well as mechanisms for galaxy quenching at very low densities.
The presence of long, straight structures in the cosmic web may serve as a test for 
cosmological models.
}
%\end{abstract}

\keywords{large-scale structure of the Universe - 
galaxies: groups: general - galaxies: clusters: general}

\maketitle

\section{Introduction} 
\label{sect:intro} 

The large-scale distribution of galaxies resembles a huge 
network called as the cosmic web. The cosmic web consists  
of galaxies, galaxy groups, and clusters connected by filaments
and separated by voids \citep{1978MNRAS.185..357J, 1988Natur.334..129K}.
Tiny density
perturbations in the very early Universe grew by 
merging and accretion of smaller structures 
\citep{1980Natur.283...47E, 1978MNRAS.183..341W,
1996Natur.380..603B}. 
The largest systems in the cosmic web 
are galaxy superclusters.
In the early studies of galaxy distribution,  superclusters
were defined as an agglomerations of galaxies and clusters, or agglomerations of clusters
('clusters of clusters'), hence the name 'superclusters'
\citep[see][ for a review of early studies of superclusters]{1983ARA&A..21..373O}.
Superclusters can be defined as the largest relatively isolated density enhancements
in the Universe \citep{1994MNRAS.269..301E, 2020A&A...637A..31S}. 
Several full-sky supercluster catalogues have been compiled using data on rich
(Abell) clusters of galaxies and the so-called friend-of-friend (percolation)
clustering algorithm \citep{1993ApJ...407..470Z, 1994MNRAS.269..301E,
2014MNRAS.445.4073C}.
This method has also been applied to determine superclusters
based on simulations \citep{2012ApJ...759L...7P}.

The advent of deep galaxy surveys made possible to calculate
the luminosity-density field of galaxies and
to determine superclusters as high-density regions in the density
field \citep{2001MNRAS.323...47B, 2003A&A...410..425E, 2003A&A...405..425E,
2004MNRAS.352..939E}.
\citet{2011MNRAS.415..964L} and \citet{2015A&A...575L..14C}
refined the definition of superclusters and proposed that superclusters 
could be defined as 
overdense regions 
that will eventually collapse in the future (so-called superstes-clusters). 
\citet{2012A&A...539A..80L} applied the luminosity-density field
to study the cosmic web and defined superclusters as
the connected high-density regions in the luminosity-density field.
The threshold density level for defining superclusters can be determined in a different way.
For example, \citet{2012A&A...539A..80L} used two criteria:
a fixed threshold density level and an adaptive density level 
which was defined for each supercluster individually, depending on
the galaxy distribution in the supercluster region. 
To determine systems which 
will collapse in the future, \citet{2011MNRAS.415..964L} 
applied a higher luminosity-density level than, for example, \citet{2012A&A...539A..80L}.
Therefore, superclusters in \citet{2012A&A...539A..80L} 
may consists  of several superstes-clusters 
\citep[see also][for examples of such superclusters]{2013MNRAS.429.3272C,
2016A&A...595A..70E}.

Superclusters embed galaxies, galaxy groups, and
clusters, connected by filaments \citep{1953AJ.....58...30D, 1958Natur.182.1478D,
1978MNRAS.185..357J, 1980Natur.283...47E}.
 As high density regions, superclusters act as great attractors, growing through the inflow of matter 
from surrounding low-density regions.
\citet{2014Natur.513...71T} analysed the cosmic velocity field 
in the nearby Universe and delineated the so-called 
basin
of attraction or the region of the dynamical influence
of the local (the Laniakea) supercluster
as a volume in space
where all galaxy flows inside it are converging. 
Examples of such regions from the local Universe were given in
\citet{2014Natur.513...71T} and \citet{2015ApJ...812...17P}
for  the Laniakea and the Arrowhead superclusters.
These authors proposed  to define superclusters on the
basis of their regions of dynamical influence. According to this definition,
a supercluster is defined as the structure inside its basin of
attraction and different superclusters are separated by the minima in the 
velocity field around them.
To keep the term 'supercluster'
tied to its conventional meaning as a high-density region in the cosmic
web, \citet{2019A&A...623A..97E} proposed calling low-density regions around
conventional superclusters supercluster cocoons.
Supercluster cocoons, together with embedded superclusters, 
correspond to the regions of
the dynamical influence, as defined by \citet{2014Natur.513...71T}.
Simulations show that supercluster cocoon boundaries follow 
the lowest density regions between superclusters \citep{2019A&A...623A..97E,
2019MNRAS.489L...1D} and the whole volume of the Universe can be divided
between supercluster cocoons.
The cocoon boundaries meet at the centre of voids, 
associated with the minima of the density field. 
Density field minima correspond to the minima  in the velocity field,
where peculiar velocities of galaxies are equal to 0 
\citep[so-called linear or the Einstein-Straus scale, ][]
{1991ApJ...379....6N, 2015A&A...577A.144T}. 
Superclusters 
occupy only approximately 1\% of the total volume, while 99\% of
the total volume is occupied by cocoons \citep{2019A&A...623A..97E}.

In this paper, we use the term 'low global density region' to denote
the volume outside supercluster  
and this definition coincides with the definition of supercluster
cocoons in general, without dividing it among superclusters. 
Low-global-density regions contain galaxy groups and  filaments. 
Rich superclusters embed high-density cores 
\citep{2007A&A...464..815E, 2007A&A...462..397E}.
Thus, it is possible to define the hierarchy of 
structures which includes high-density cores of superclusters,
their lower density outskirts regions, and low-global-density cocoon regions around them,
which are taken altogether as basins in the definition from \citet{2014Natur.513...71T}.

We note that sometimes the term 'field' is used.
In cluster studies, 'field' may denote regions outside galaxy clusters
(even regions between clusters inside superclusters). 
In some studies, the term 'field' have been used to denote regions outside
superclusters, the same as 'low global density regions' or 'cocoons' in the present study.
To avoid misunderstandings, 
we do not use the term 'field' in our study.

The essential evolution of superclusters and their
components occurs inside supercluster cocoons 
\citep{2019A&A...623A..97E}.
Simulations show that present-day rich clusters 
have collected their  galaxies along filaments from regions with comoving radii 
at least $10$~\Mpc\ at redshift $z = 1$
\citep{2013ApJ...779..127C, 2016A&ARv..24...14O}. The formation
of clusters is accompanied by the star-formation quenching
of galaxies when they fall into clusters \citep{2015ApJ...806..101H, 2018A&A...620A.149E,
2018MNRAS.476.4877M, 2019MNRAS.483.3227K, 2019A&A...621A.131M, 2020MNRAS.491.5406T}. 
One result of galaxy quenching and morphological transformations is the large-scale
morphological segregation of galaxies 
at low and high redshifts.
According to the large-scale
morphological segregation galaxies in high-global-density regions (clusters and superclusters) 
are of early type, 
red, and passive. In low-global-density regions galaxies are 
mostly of late type, blue, and star forming 
\citep{1974Natur.252..111E, 1980ApJ...236..351D, 
1986ApJ...300...77G, 1987MNRAS.226..543E, 2003MNRAS.346..601G, 2003ApJ...584..210G,
2004MNRAS.353..713K, 2004ApJ...615L.101B, 2006A&A...458...39C,
2007A&A...470..425B, 2007A&A...464..815E, 2007A&A...468...33E, 
2011A&A...529A..53T, 2012ApJ...746..188M, 2017ApJ...835..153F,
2017MNRAS.465.3817M}.
However, red galaxies with old stellar populations can 
also be found in poor galaxy groups in low-global-density regions  
\citep{2007ApJ...658..898P, 2012A&A...545A.104L}.

Therefore, star formation quenching occurs
also in poor groups and filaments in both high and low global density environments. 
If this happens before their infall to clusters, then
this is called preprocessing
\citep{2009MNRAS.400..937M,
2017A&A...607A.131D, 2018A&A...610A..82E, 2018MNRAS.473L..79B, 2018MNRAS.475.4148G,
2019A&A...632A..49S, 2019MNRAS.484.4695T}. 
Also, star-forming galaxies with young stellar populations
have been observed in clusters and in their infalling structures, 
even  in the high-density cores
of superclusters \citep{2010A&A...522A..92E, 2016MNRAS.461.1202J, 2018A&A...610A..82E}.
Thus, the details of galaxy evolution and star formation 
quenching, as well as the physical processes
which shape galaxies in various environments, are not yet clear.

The star-formation quenching processes due to removal of gas can be divided 
as external and internal ones. 
A comprehensive overview of different physical processes 
triggering removal of gas in galaxies and their relative 
importance is given by \citet{2006PASP..118..517B}.

Internal processes, called also as mass quenching, depend, first of all, on galaxy dark halo mass. 
Such internal processes blowing out galactic gas include stellar winds, 
supernovae explosions, nuclear activity, and so on
 \citep{2006MNRAS.372..265M, 2006MNRAS.365...11C,
2019MNRAS.485.3446H}.
Mass quenching is more effective for massive galaxies 
 \citep{2020ApJ...889..156C} and at higher redshifts.

External processes or environmental quenching are due 
to stripping away galactic gas by the ram pressure from 
a cluster's or group's hot gas 
\citep{1972ApJ...176....1G, 2008MNRAS.383..593M,
2019ApJ...873...42R, 2019MNRAS.483.1042Y},
cold gas removal by viscous stripping 
\citep{1982MNRAS.198.1007N}, 
starvation due to the prevention of fresh gas from reaching galaxies by removal their 
feeding primordial filaments 
\citep{2019OJAp....2E...7A, 2019A&A...621A.131M},
harassment due to multiple high-speed mergers 
\citep{1996Natur.379..613M}. 
External processes depend on environmental densities, 
on galaxy orbital properties, and also on galaxy masses. 
They are most effective at lower redshifts and for intermediate mass galaxies 
\citep{2020ApJ...889..156C}. 
Tidal interactions between galaxies are external processes and they can 
transform the morphology of galaxies significantly. 
However, in the densest parts of galaxy clusters, the relative speed of galaxies 
is quite high and these interactions do not affect the SF significantly 
\citep{2006PASP..118..517B}.

The relative importance of these mechanisms is not clear as corresponding 
physical processes depend on several unknown (free) parameters. 
Often, several mechanisms work together 
\citep{2015Natur.521..192P, 2010ApJ...721..193P, 2020MNRAS.492...96B}; 
for example, \citet{2020MNRAS.491.5406T} derived that 
for most of galaxies in their study, starvation, along with 
certain contributions from outflow, best fits the observed galactic properties.
Still, even processes considered 'global' in these studies are related to 
the local environment of galaxies in the context of our paper
and the question of how very large scales affect the properties 
of galaxies is still open.

A detailed study of the distribution and properties of
galaxies and their systems in superclusters and in low global density regions
(cocoons) around superclusters helps us  
to clarify how the local and global
environmental effects combine in determining the evolution
and present-day dynamical state of superclusters and their components,
along with the star-formation history of galaxies in these structures.
Superclusters and their cocoons, together with the large variety of 
environments, ranging from rich clusters in high-density cores of superclusters
to lowest global density regions with poor groups at cocoon boundaries, serve as good laboratories
for such studies.

In this paper, we focus on the study 
and comparison of the supercluster SCl~A2142 and the cocoon region around it. 
SCl~A2142 is named after the richest galaxy cluster 
in it, which is the Abell cluster \object{A2142}. It is located at redshift $z \approx 0.09$
(at distance $264.5$~\Mpc).
The length of the supercluster
is $\approx 50$~\Mpc, and its mass is $M \approx 4.3\times~10^{15}h^{-1}M_\odot$
\citep{2012A&A...539A..80L}.
SCl~2142 has an almost spherical main body with a 
radius approximately $13$~\Mpc\ which embeds the HDC with radius of
about $5$~\Mpc, and a straight tail 
populated by galaxy groups and short, thin filaments \citep{2015A&A...580A..69E}.
The tail as a whole resembles the thick filament 
described in \citet{2013MNRAS.429.1286C}. 
Earlier studies have shown that the supercluster main body is 
already collapsing or  will collapse in the future \citep{2015A&A...580A..69E, 
2015A&A...581A.135G, 2018A&A...620A.149E}. 

Our goal in this study is to understand the growth and possible future evolution
of the supercluster SCl~A2142 and its components, along with the galaxy transformations taking place within it. 
With this purpose in mind, we determine the boundaries of SCl~A2142 cocoon
and analyse the properties of galaxies and galaxy systems 
in it. Our study extends the previous analysis of the supercluster SCl~A2142
to its environment up to the lowest global-density regions between superclusters
\citep{2015A&A...580A..69E, 2018A&A...610A..82E, 2018A&A...620A.149E}.
We aim to connect the properties of supercluster components  with
the dynamical state and main evolutional epochs  of the supercluster.
SCl~A2142, with its rather simple structure, is a good object for such studies.
It is surrounded by low-global-density regions and its nearest rich supercluster 
is the Corona Borealis supercluster, located at a distance of approximately $40$~\Mpc\
from SCl~A2142 \citep{2015A&A...580A..69E}.
This is the first study with this type of analysis for a supercluster
and its environment. 

We determined the boundaries of the supercluster cocoon by an  
analysis of the luminosity-density field around the supercluster.
We located filaments and groups in the cocoon and looked for their 
connectivity in and around the supercluster. We also studied the connectivity of the
HDC of the supercluster,
and its central cluster, A2142.
The connectivity is a parameter defined 
as the number of filaments connected to a cluster  \citep{2000PhRvL..85.5515C,
2018MNRAS.479..973C}. 
The connectivity characterises  the growth of cosmic structures and 
is related to the properties of the dark matter and dark energy.
Connectivity can be predicted theoretically; 
\citet{2018MNRAS.479..973C} showed that on average, cluster connectivity
depends on cluster mass and richness.
Connectivity of poor groups is lower than that 
of rich groups \citep[see also][]{2019MNRAS.489.5695D,
2020MNRAS.491.4294K, 2020A&A...635A.195G}.

To understand the evolution of groups and galaxies in them, 
we compared galaxy populations and group properties in the supercluster 
and in the low-global-density regions around it.
To assess the evolutionary state of galaxy groups, we used the magnitude
gap between the brightest galaxies in them as an indicator
of their evolutionary status \citep[][and references therein]{2018MNRAS.474..866V}.
\citet{2018MNRAS.474..866V} showed that groups with large magnitude gaps
between their brightest galaxies tend to be more concentrated,
supporting the possibility that they may have been formed earlier. They 
are also more relaxed, as has been found in simulations \citep{2013ApJ...777..154D,
2014A&A...566A.140G, 2016ApJ...824..140R}.

Among the variety of galaxy properties we focussed on their
star formation and looked for galaxies at various epochs of their star-formation
history, starting from  blue star-forming galaxies.
Our main focus is to study galaxies in transformation, as 
red star forming galaxies and recently quenched galaxies. 
This  study may provide observational evidence
about the cosmic web detachments \citep{2019OJAp....2E...7A}.
According to this phenomenon, a galaxy is quenched when accretion of
gas along primordial filaments is halted.
We also analysed the distribution of  galaxies with very old stellar populations,
which have been found in somewhat extreme high-density environments of centres of 
galaxy clusters.
Therefore, we expect that such galaxies could be rare in 
poor groups in low-global-density environments.

In this paper, we use, as in \citet{2018A&A...620A.149E}, 
data from the Sloan Digital Sky Survey data release 10
(SDSS DR10) to analyse the structure and properties of galaxies and galaxy
systems in the cocoon of SCl~A2142 \citep{2011ApJS..193...29A, 2014ApJS..211...17A}.
In accordance with the earlier studies of the A2142 supercluster, 
we use the following cosmological parameters: the Hubble parameter $H_0=100~ 
h$ km~s$^{-1}$ Mpc$^{-1}$, matter density $\Omega_{\rm m} = 0.27$, and 
dark energy density $\Omega_{\Lambda} = 0.73$ 
\citep{2011ApJS..192...18K}.

\section{Data} 
\label{sect:data} 

\subsection{Supercluster, group, and filament data}
\label{sect:gr}

\begin{table*}[ht]
\caption{Data on Scl~A2142.}
\label{tab:a2142}  
\begin{tabular}{rrrrrrrrr} 
\hline\hline 
\multicolumn{1}{c}{(1)}&(2)&(3)&(4)& (5)&(6)&(7)&(8)&(9)\\      
\hline 
\multicolumn{1}{c}{ID}& $N_{\mathrm{gal}}$& $\mathrm{R.A.}$ & $\mathrm{Dec.}$ & $\mathrm{Dist.}$  
 & $L$ & $M$ & $\mathrm{Length}$ & $D8_{\mathrm{max}}$ \\
& &[deg]&[deg]&[$h^{-1}$ Mpc]&[$10^{13}h^{-2} L_{\sun}$]&[$10^{15}h^{-1}M_\odot$]& [$h^{-1}$ Mpc] & \\
\hline
 239+027+009 & 1038  & 239.52 & 27.32 & 264.5&  1.6 & 4.3 &  50.3 & 21.6 \\ 
\hline
\end{tabular}\\
\tablefoot{                                                                                 
Columns in the Table are as follows:
(1): Supercluster ID AAA+BBB+ZZZ, where AAA is R.A., +/-BBB is Dec., and ZZZ is 100$z$;
(2):  Number of galaxies in the supercluster, $N_{\mathrm{gal}}$;
(3--5): Supercluster centre right ascension, declination, and comoving distance
at the cluster A2142;
(6): Luminosity of the supercluster, $L$;
(7):  mass of the supercluster, $M$;
(8):  Supercluster length (the maximum distance between galaxies in
the supercluster), $\mathrm{Length}$;
(9):  Maximal value of the luminosity-density field calculated with
the $8$~\Mpc\ smoothing kernel, $D8_{\mathrm{max}}$, in units of the mean density as described in the text.
}
\end{table*}

\begin{table*}[ht]
\caption{Data on galaxy groups with at least ten member galaxies
in the HDC of SCl~A2142, and in the SCl~A2142 cocoon.}
\begin{tabular}{rrrrrrrrrrrrrrr} 
\hline\hline  
(1)&(2)&(3)&(4)&(5)& (6)&(7)&(8)&(9)&(10)&(11)&(12)&(13)& (14) & (15)\\      
\hline 
No. & ID&$N_{\mathrm{gal}}$& $\mathrm{R.A.}$ & $Dec.$ 
&$\mathrm{Dist.}$ &$\mathrm{D_C}$ &  $R_{\mathrm{vir}}$ & $L_{\mathrm{tot}}$  
& $M_{\mathrm{dyn}}$ & $D1$& $D8$ & $|\Delta M_{12}|$ & $dV$ & $\pazocal{C}$\\
\hline                                                    
 1 & 10570 &  27  & 238.53 & 27.47 &  268.6 & 4.4  & 0.53 & 51.3 &  0.6& 245.2 & 13.7 & 0.04  & 236 & 0  \\
 2 &  3070 & 212  & 239.52 & 27.32 &  264.6 & 0.0  & 0.88 &382.0 &  9.1& 600.0 & 20.7 & 1.24  & 334 & 6-7\\
 3 &  4952 &  54  & 239.78 & 26.56 &  260.1 & 3.3  & 0.70 &111.0 &  2.1& 418.8 & 17.1 & 0.33  & 149 & 1\\
 4 & 32074 &  11  & 240.11 & 26.71 &  262.4 & 3.3  & 0.28 & 15.7 &  0.6& 164.4 & 19.9 & 0.17  & 284 & 1\\
 5 & 35107 &  10  & 240.13 & 27.01 &  258.7 & 2.4  & 0.47 & 13.9 &  0.3& 110.8 & 14.2 & 0.18  & 387 & 1\\
 6 & 14960 &  27  & 240.20 & 25.87 &  263.6 & 7.0  & 0.59 & 46.6 &  1.5& 209.0 & 15.3 & 0.58  & 607 & 1\\
 7 & 17779 &  20  & 240.38 & 26.16 &  261.1 & 6.0  & 0.40 & 35.1 &  1.0& 279.4 & 16.4 & 0.0   & 566 & 1\\
 9 & 21183 &  21  & 240.83 & 26.95 &  265.0 & 5.4  & 0.34 & 36.5 &  0.6& 248.0 & 15.2 & 0.04  & 293 & 1\\
\hline
C1 & 13822 &  10 & 239.92 & 21.03 &  261.6 & 29 & 0.27 & 16.0  & 0.1  & 171.3 & 4.4 & 1.28  & 113 & 0 \\
C2 & 20637 &  10 & 241.03 & 28.92 &  257.5 & 10 & 0.42 & 14.8  & 0.4  & 111.0 & 3.7 & 0.59  & 198 & 0 \\
C3 &  9352 &  13 & 242.87 & 29.54 &  271.4 & 18 & 0.36 & 27.0 &  0.3  & 289.5 & 3.7 & 0.85  & 198 & 1 \\
C4& 14236  &  11 & 236.46 & 27.49 &  273.0 & 15 & 0.21 & 30.0 &   0.4 & 374.6 & 3.3 & 0.04  & 188  & 1\\
\hline                                                                                       
$G_{gap}$ &  7481 &  8 & 239.62 & 21.58 &  263.2 & 26 & 0.27 & 16.9 & 0.2 &220& 3.4 & 1.45  & 113  & 1\\
$G_{VO}$ & 13360 &   7 & 246.07 & 23.57 &  271.8 & 32 & 0.25 & 19.4 & 0.2 &205& 1.8 & 0.80  &  90  & 1 \\
%$G_{RQ}$ &  8992 &  24 & 242.95 & 16.97 &  264.7 & 50 & 0.38 & 43.1 & 1.3 & 3.8 & 1.64  &  67  & 1\\
\hline
\hline
\label{tab:gr10}  
\end{tabular}\\
\tablefoot{                                                                                 
Columns are as follows:
(1): Order number or ID of the group (see text);
(2): ID of the group from \citet{2014A&A...566A...1T} (Gr~3070
correspond to the Abell cluster A2142);
(3): Number of galaxies in the group, $N_{\mathrm{gal}}$;
(4)--(5): Group centre right ascension and declination (in degrees);
(6): Group centre comoving distance (in $h^{-1}$ Mpc);
(7): Group distance from the  centre of the cluster A2142
(for brevity, clustercentric distance) (in $h^{-1}$ Mpc);
(8): Group virial radius (in $h^{-1}$ Mpc);
(9): Group total luminosity (in $10^{10} h^{-2} L_{\sun}$);
(10): Dynamical mass of the group assuming the NFW density profile, $M_{\mathrm{dyn}}$,
(in $10^{14}h^{-1}M_\odot$);
(11-12): Luminosity-density field values at the location of the group, $D1$ and $D8$, 
in units of the mean density as described in the text.
(13): Magnitude gap between the two brightest galaxies in a group.
(14): Difference between the velocity of the main galaxy in a group, and mean group
velocity (in $km s^{-1}$).
(15): Connectivity $\pazocal{C}$ (the number of filaments or subsystems connected to 
a group). 
}
\end{table*}

We used data from supercluster, group, and filament catalogues by 
\citet{2012A&A...539A..80L} and \citet{2012A&A...540A.106T, 2014A&A...566A...1T,
2014MNRAS.438.3465T, 2016A&C....16...17T}, based on the SDSS MAIN galaxy dataset.
Catalogues of galaxy superclusters, groups, and filaments are
available from the database of cosmology-related catalogues 
at \url{http://cosmodb.to.ee/} , where the corresponding catalogues can be found.
In compiling the catalogues used in this paper, we used the SDSS DR10 MAIN spectroscopic galaxy sample  with the 
apparent Galactic extinction corrected $r$ magnitudes $r \leq 
17.77,$ and redshifts of $0.009 \leq z \leq 0.200. $

\subsection{Galaxy data}
\label{sect:galpop}

Data on galaxy properties used in this study are taken 
from the SDSS DR10 web page  
\footnote{\url{http://skyserver.sdss3.org/dr10/en/help/browser/browser.aspx}}.
Absolute magnitudes of galaxies have been calculated as
\begin{equation}
M_r = m_r - 25 -5\log_{10}(d_L)-K,
\end{equation} 
where $d_L$ is the luminosity distance in units of $h^{-1}$Mpc and
$K$ is the $k$+$e$-correction, calculated as in 
\citep{2007AJ....133..734B} and  \citet{2003ApJ...592..819B}
\citep[see][for details]{2014A&A...566A...1T}.
The distance to the SCl~A2142 is $ \approx 265$~\Mpc;
at this distance the galaxy sample is complete
at the absolute magnitude limit $M_r = -19.6$ in units of $\mathrm{mag}+5\log_{10}h$. 
We use the full sample of galaxies in SCl~A2142 environment
to determine the boundaries of the supercluster cocoon and to detect
galaxy groups and filaments within it. For a statistical comparison
of the galaxy content of filaments and groups, we use
 a magnitude-limited complete sample,  with the magnitude limit $M_r = -19.6$~mag.

To calculate galaxy colours and the concentration index, 
we use galaxy magnitudes and Petrosian radii from the SDSS photometric data.
The galaxy rest-frame colour index $(g - r)_0$ is defined as $(g - r)_0 = M_g - M_r$,
where $M_g$ and $M_r$ are absolute magnitudes of galaxies in $g$ and $r$ band. 
The concentration index of galaxies $C$ is calculated as the ratio of the 
Petrosian radii $R_{50}$ and  $R_{90}$: $C = R_{50}/R_{90}$;
Petrosian radii $R_{50}$ and  $R_{90}$ are defined as radii which contain
50\% and 90\% of the Petrosian flux of a galaxy \citep{2001AJ....121.2358B,
2001AJ....122.1104Y}.
The concentration index is related to the galaxy structure parameters 
\citep{2005AJ....130.1535G}
and {more weakly, to the galaxy morphological type. 

The probability of being an early-type or late-type galaxy 
is taken from \citet{2011A&A...525A.157H}.
\citet{2011A&A...525A.157H} used a Bayesian automated 
morphological classification  of galaxies, which assigned to each galaxy 
a probability of being of early or late type. They divided galaxies into four 
main classes (E, S0, Sab, Scd), and calculated for each galaxy, the probabilities
$P_E$, $P_{S0}$, $P_{Sab}$, and $P_{Scd}$. In our analysis, we consider the probabilities of Sab or Scd type together as a probability to be a late-type galaxy;
$P_{late} = P_{Sab} + P_{Scd}$. A galaxy can be considered of late type if
$P_{late} \geq 0.5$.

\begin{figure}[ht]
\centering
\resizebox{0.46\textwidth}{!}{\includegraphics[angle=0]{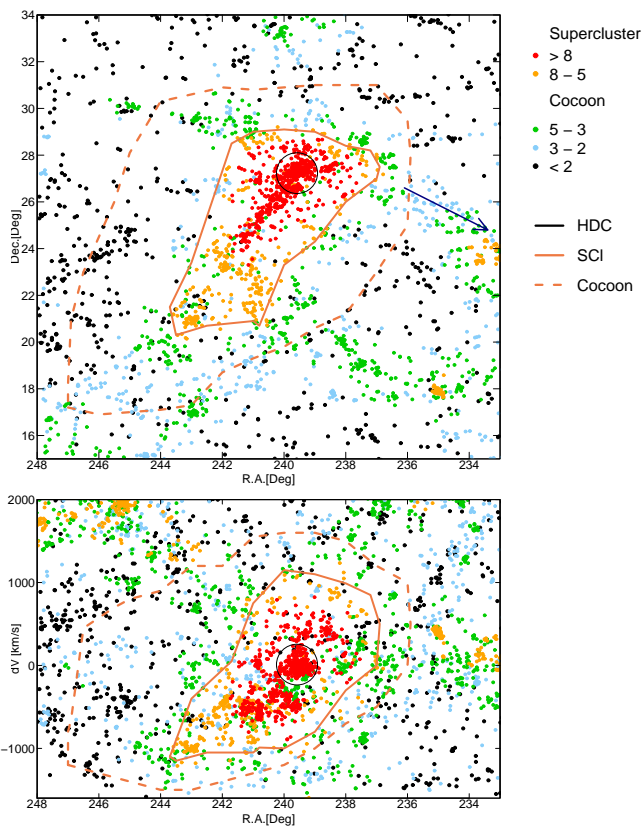}}
\caption{
Distribution of galaxies in the supercluster SCl~A2142 region. Upper panel shows
galaxy distribution in the sky plane,
and lower panel  in the R.A. versus the difference between the galaxy velocity and 
the velocity of the cluster A2142 centre ($dV$). In upper panel distance  limits are $250 - 275$~\Mpc\
(the distance to the A2142 centre is $\approx 265$~\Mpc),
and in lower panel declination limits are $17 - 31$~degrees.
Colours denote galaxies in regions
of different luminosity-density as follows. 
Red: $D8 \geq 8$ (high-density main body of the supercluster), 
yellow: $5 < D8 \leq 8$ (outskirts of the supercluster), green: $3 < D8 \leq 5$,
blue: $2 < D8 \leq 3$, and black: $D8 \leq 2$, where $D8$ denotes
the luminosity-density in units of the mean luminosity-density.
Orange solid line approximately marks the boundaries of the supercluster, and
orange dashed line shows the cocoon boundaries. Black circle
shows the HDC of the supercluster. 
Navy arrow points towards the Corona Borealis supercluster.
}
\label{fig:radecvel}
\end{figure}

We use data about stellar masses $M^{\mathrm{*}}$,  star formation rates ($SFR$),
metallicities $Z$,
and $D_n(4000)$ index of galaxies from 
the MPA-JHU spectroscopic catalogue \citep{2004ApJ...613..898T, 2004MNRAS.351.1151B}.  
In this catalogue, the properties of 
galaxies are calculated using 
the stellar population synthesis models and fitting SDSS photometry and spectra 
with \citet{2003MNRAS.344.1000B} models.
The description of how the stellar masses of galaxies are calculated 
can be found in \citet{2003MNRAS.341...33K}.
The  $D_n(4000)$ index
(the ratio of the average flux densities
in the band $4000 - 4100 \angstrom$ and $3850 - 3950 \angstrom$)
is defined as in \citet{1999ApJ...527...54B}. 
The $D_n(4000)$ index is correlated with the time passed 
from the most recent star formation event in a galaxy.
It can be used as proxy for the age of stellar populations of galaxies,
and star formation rates.
We also used stellar velocity dispersions of galaxies, $\sigma^{\mathrm{*}}$. 
They are from Portsmouth group, and calculated by fitting galaxy spectra 
using publicly available codes, namely the Penalized PiXel Fitting 
\citep[pPXF,][]{2004PASP..116..138C} 
and Gas and Absorption Line Fitting code 
\citep[GANDALF,][]{2006MNRAS.366.1151S}.

\section{Luminosity-density field and the boundaries of the SCl~A2142 cocoon}
\label{sect:cocoon}

As the first step to determining superclusters in the SDSS MAIN dataset, 
the global luminosity-density field was calculated using
$B_3$ spline kernel with the smoothing length 8~\Mpc:
\begin{equation}
    B_3(x) = \frac{1}{12} \left(|x-2|^3 - 4|x-1|^3 + 6|x|^3 - 4|x+1|^3 + |x+2|^3\right).
\end{equation}
Superclusters have been defined as the connected 
volumes above a threshold density of $D8 = 5.0$
(in units of mean luminosity-density, $\ell_{\mathrm{mean}}$ = 
1.65$\cdot10^{-2}$ $\frac{10^{10} h^{-2} L_\odot}{(\vmh)^3}$,
in the luminosity-density field). 
The use of smaller smoothing length gives high-density regions which
correspond to clusters of galaxies, or to high-density cores of superclusters,
depending on the smoothing length value, from $1$ to $4$~\Mpc.
Very large smoothing length means that the largest connected regions have 
much larger sizes than conventionally correspond to individual
superclusters
\citep[see, for example, ][]{2018A&A...616A.141E, 2020arXiv200503480E}.
Also, if we change the density level used to define individual superclusters,
then choosing a higher density level, has the same effect as choosing as
a small smoothing length which selects high density cores of superclusters.
A choice of a too low density level means that we may combine together 
neighbouring superclusters,
as  shown in the case of the Sloan Great Wall in \citet{2011ApJ...736...51E}.
Details about the luminosity-density field and 
supercluster definition can be found in   \citet{2012A&A...539A..80L}.

We used the luminosity-density field to determine the 
supercluster SCl~A2142 and to find its cocoon boundaries. 
The supercluster centre is at the rich galaxy cluster A2142.
The supercluster is defined as a volume around the cluster A2142 with luminosity-densities of
$D8 \geq 5.0$. With its redshift at $z \approx 0.09,$
the supercluster is located well within the SDSS MAIN galaxy sample.
The luminosity-density drops at the borders of the supercluster and,
thus, 
SCl~A2142 defined using a fixed luminosity limit,
and with an adaptive density limit coincide.
We show the basic data on the supercluster SCl~A2142 in Table~\ref{tab:a2142}, which
provides the centre coordinates of the supercluster at the cluster A2142,
the number of galaxies, and the luminosity, mass, size, and maximum 
global density $D8_{\mathrm{max}}$ in it.

\begin{figure}[ht]
\centering
\resizebox{0.44\textwidth}{!}{\includegraphics[angle=0]{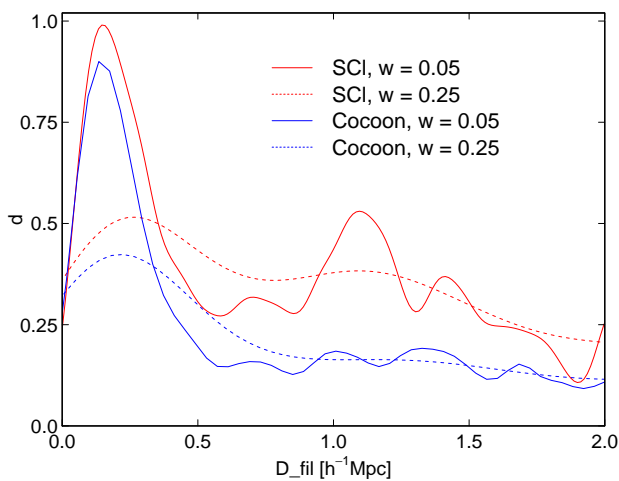}}
\caption{
Distribution of distances $D_{fil}$ from the nearest filament axis 
for galaxies in the supercluster (SCl, red lines) and in the cocoon
(cocoon, blue lines).
Solid lines show distributions calculated with high resolution kernel
with width $w = 0.05$, and dashed lines show distributions calculated with
low resolution kernel width, $w = 0.25$.
}
\label{fig:dfil}
\end{figure}

\begin{figure}[ht]
\centering
\resizebox{0.47\textwidth}{!}{\includegraphics[angle=0]{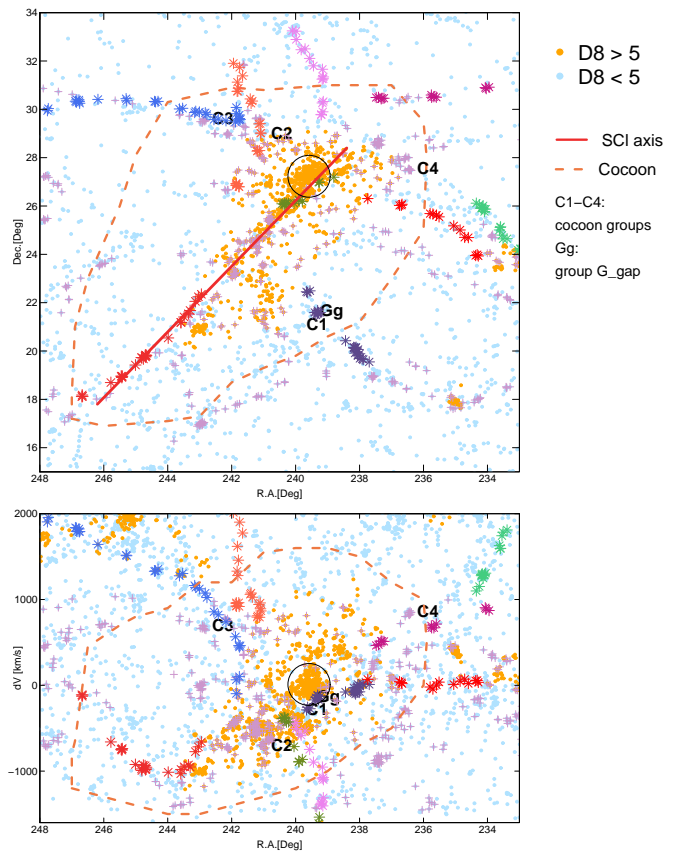}}
\caption{
Distribution of galaxies, 
groups,  and filaments
around the supercluster SCl~A2142 in the sky plane (upper panel)
and in the R.A. - velocity difference plane (lower panel). 
Orange dots  denote galaxies 
in the supercluster ($D8 > 5$) and light blue dots mark galaxies 
in low global density region ($D8 < 5$). 
Member galaxies of individual filaments
with length $\geq 20$~\Mpc\ are denoted with stars of different colours and
galaxies in filaments shorter than $20$~\Mpc\ with pale violet crosses. 
Dark red line shows supercluster axis and filament F1033 at its extension. 
Numbers show ID numbers of groups with at least ten member galaxies
in the cocoon (C1-C4, Table~\ref{tab:gr10}), and 
group $G_{gap}$ is denoted as $Gg$ (see text).
Orange dashed line shows the cocoon boundaries, and black circle shows 
the HDC boundaries.
}
\label{fig:radecvelfil}
\end{figure}

\citet{2011A&A...532A...5E, 2015A&A...580A..69E} and \citet{2015A&A...581A.135G} 
studied the morphology of SCl~A2142 and luminosity-density distribution within it
and showed that the supercluster has  a HDC where the luminosity-density is 
$D8 > 17$. The radius of the HDC is approximately $5$~\Mpc. It is embedded in 
 an almost spherical main body of the supercluster with the 
radius of about $13$~\Mpc. SCl~A2142 also has an almost straight tail,
so that the total length of the supercluster is approximately $50$~\Mpc.  
The outer parts of the main body and its tail with the luminosity-density of 
$8 > D8 \geq 5$ form the outskirts region of the supercluster.
Below in the text we refer to the 'supercluster' when we are analysing the whole
supercluster. We refer to the HDC or to the outskirts regions when these are being analysed individually.

We show the distribution of galaxies in SCl~A2142 and in its large-scale environment
in Fig.~\ref{fig:radecvel}.
Upper panel shows the distribution of galaxies in the sky plane.
In the lower panel, the $x$-axis is the right ascension and the $y$-axis
shows the line-of-sight velocity difference between a galaxy and the supercluster
centre at the cluster A2142, calculated using redshifts of galaxies. 
The velocities of galaxies in groups
are corrected for the Fingers-of-God effect, as described in 
\citet{2012A&A...540A.106T, 2014A&A...566A...1T}.

Cocoon boundaries are defined as minima in the density field between  superclusters.
In  Fig.~\ref{fig:radecvel}, the location of galaxies is colour-coded according to the
value of the global luminosity-density at their locations. 
In what follows, we refer to the region with a luminosity-density of
$D8 \geq 5$ as the supercluster SCl~A2142; this density limit determines
the boundary between the supercluster and cocoon. The region with lower   global luminosity-density represents the cocoon. Black dots, which show the lowest  global luminosity-density,  with $D8 \leq 2,$ approximately delineate the outer boundary of SCl~A2142
cocoon. In fact, the density threshold for cocoon borders is not fixed.
It is defined as a local minimum in the density distribution
and may vary in different locations. Cocoon boundaries can be
followed by orange line in Fig.~\ref{fig:radecvel}. The average distance between
the supercluster and its cocoon boundaries is approximately $10 - 15$~\Mpc.
This limit separates the SCl~A2142, the Corona Borealis
supercluster (its direction is shown with arrow in Fig.~\ref{fig:radecvel})
and other superclusters close to SCl~A2142.
Figure~\ref{fig:radecvel} shows that there are groups with 
$D8 > 2$ outside SCl~A2142 cocoon boundaries, for example, in the direction
of the  Corona Borealis supercluster, indicated by the arrow. These groups belong
to the cocoon of the Corona Borealis supercluster. 

The exact determination of SCl~A2142 cocoon boundaries is not straightforward
since we do not know the distribution and velocities of galaxies in low global density
environment of SCl~A2142 in detail. 
In our calculations,  
we use for simplicity, following sky coordinate and
distance limits as SCl~A2142 cocoon boundaries: right ascension $236 \leq R.A. \leq 246.1$~degrees, declination 
$17 \leq Dec. \leq 31$~degrees, and distance $250 \leq D \leq 275$~\Mpc\
($0.084 \leq z \leq 0.10$).
We note that although it is difficult 
to define cocoon boundaries with density field data only
\citep[as also discussed by ][]{2019MNRAS.489L...1D}, we use this approach since
there is no peculiar velocity data for SCl~A2142 region yet (nor coming soon). 
Both supercluster and cocoon sample have 
$\approx 1000$ galaxies. 

In our analysis of the local environment of galaxies, we also use
local luminosity-density values calculated with the smoothing length $1$~\Mpc\
($D1$).
This scale corresponds to the sizes of galaxy groups and dark matter haloes
of galaxies \citep[see][for references]{2018A&A...616A.141E}.

\section{Filaments and groups in the supercluster and in the cocoon }
\label{sect:fil}

{\bf Galaxy filaments} were detected by 
applying marked point process to the SDSS galaxy distribution (Bisous model)
 \citep{2014MNRAS.438.3465T, 2016A&C....16...17T}. 
For each galaxy, a distance from the nearest filament
axis was calculated.  
In 
Fig.~\ref{fig:dfil}, we show for each galaxy in the supercluster and in the cocoon,
the distribution of their distances from the nearest filament axis ($D_{fil}$). 
In this figure, we do not show filament distances for galaxies from the HDC
and we discuss the structures in the HDC separately.
Figure~\ref{fig:dfil} shows that at the distance from filament axis $D_{fil} > 0.5$~\Mpc,\
the number of galaxies drops rapidly in both environments. 
Therefore, galaxies are considered as filament members 
if their distance from the nearest filament axis was within $0.5$~\Mpc\
\citep[see][for details]{2014MNRAS.438.3465T}.
Galaxies which belong to a group may  also be members of a filament. 

As seen in Fig.~\ref{fig:dfil}, the number of galaxies with filament distances
in a range of $\approx 0.5 - 1.0$~\Mpc\ is very small, which means that of we  
 use slightly larger $D_{fil}$ limit (for example,
$0.8$~\Mpc), then only a few galaxies are added to filaments.  
In the supercluster, at $D_{fil} \approx 1.0,$ there is a secondary maximum
in the distance distribution, indicating that neighbouring filaments are located
closely together. There is no such peak in the filament distance distribution
in the cocoon, showing that in the low-global-density environment, mutual distances
between individual systems
are larger than in the supercluster, which could be expected.

Another observation based on this figure is that by using
different kernel widths, $w,$ to calculate distance distributions,
large values of $w$ mimic the case if the filament finder is fine-tuned so
that it detects wide filaments, with widths up to a few megaparsecs.
In this case, all the details of distributions get lost,
especially in the high-global-density environment of the supercluster where the use of wide
filaments may combine most of the structures. 
However, in a low-global-density environment (approximately
99\% of the total volume, as mentioned in Sect.~\ref{sect:intro}), where most of the filaments lie,
it may be even preferable method to trace individual filaments, as more galaxies become
filament members. 

Moreover, filaments detected using different methods or with different criteria for
filament membership
may have different properties.
The Bisous method used to determine filaments
in \citet{2014MNRAS.438.3465T}, used in this study,
tends to find thinner filaments than, for example,
in the Disperse method
\citep{2011MNRAS.414..350S} or Nexus method \citep{2013MNRAS.429.1286C}.
This is shown, for example, in Fig. 16 by \citet{2020A&A...637A..18B},
and in Fig. 4 by \citet{2019MNRAS.487.1607G}.
\citet{2020arXiv200201486M} showed that the properties of filaments
depend on the density of a galaxy sample used
to determine filaments. Also, short filaments, especially near
clusters or within superclusters, as the ones in SCl~A2142 described below,
may be  difficult to detect
and are often excluded from the sample \citep{2020arXiv200201486M, 2020A&A...637A..31S,
2020arXiv200604463K}.

In addition, Fig.~\ref{fig:dfil} shows that using wide filaments (large value of 
$D_{fil}$ for filament membership) may
result in merging close thin filaments together in high-global-density environments, decreasing the connectivity
of a group or cluster. An example of this are structures with small mutual distances
connected to the cluster A2142 in the HDC of SCl~A2142 (Fig.~\ref{fig:hdcradec}).
An extreme example of the use of wide filaments with 
$D_{fil}$ values $D_{fil} > 2$~\Mpc\ is that in this case all thin filaments
along SCl~A2142 axis may become members of one thick filament. 
This would lead to a loss of all the information about the 
structure of this part of the supercluster.
Moreover, increasing $D_{fil}$ values may also connect groups to filaments,
increasing the number of connected groups, or
increasing the connectivity of already connected groups. 
In contrast, in low-global-density environments, using thick filaments may not increase
the connectivity of individual groups but it may increase
the number of groups connected to filaments. These examples
show that the comprehensive analysis of various
possibilities 
deserves a special study, which is beyond the scope of the present paper.

Most filaments detected in the cocoon of SCl~A2142 have a length of less than $10$~\Mpc.
We detected 70 such filaments in the cocoon, and 23 filaments with
length between $10$ and $20$~\Mpc.
There are also seven long filaments with lengths over $20$~\Mpc\  
in the cocoon. One of them, F1033, starts in the supercluster
tail.
For comparison, \citet{2015A&A...580A..69E} detected 31 filaments 
in SCl~A2142. Of these filaments, 27 were shorter than $10$~\Mpc,
and 4 were of a length between $10$ and $20$~\Mpc.
In Fig.~\ref{fig:radecvelfil} we show the distribution of galaxies 
in the sky plane and in the sky-velocity difference plane. In this figure we mark galaxies which belong
to filaments of different lengths. 
The most interesting among long filaments which surround the supercluster
is a filament with ID number F1033 and 
length of $36$~\Mpc\ in the catalogue. This filament  has an almost straight shape
in the sky plane (except the farthest galaxy at the
end of the filament, at $R.A > 246$; the velocity of this  galaxy is also
different from velocities of other galaxies in this filament). 
It is located along the supercluster axis and begins in
the supercluster tail.
Supercluster axis with F1033 is shown with a dark red line in Fig.~\ref{fig:radecvelfil}.
The total length of SCl~A2142 together with this filament
is approximately $75$~\Mpc, which makes it the longest straight structure
in the Universe described so far.

To analyse the {\bf group} content of the supercluster environment,
we use the catalogue of galaxy groups by \citet{2014A&A...566A...1T}.
This catalogue was used to find groups in the supercluster also by \citep{2018A&A...620A.149E}.
\citet{2014A&A...566A...1T} applied a friends-of-friends (FoF) cluster analysis 
method to determine groups in the galaxy distribution. According to FoF, a 
galaxy belongs to a 
group if this galaxy has at least one group member galaxy closer 
than a linking length. A description of the data reduction and the choice of 
the linking length are given in \citet{2014A&A...566A...1T}.

In the A2142 cocoon, there are 280 galaxy groups, 
4  of them having 10 - 13 member galaxies.
In contrast,
in the supercluster 14 groups have at least 10 member galaxies,
and 10 of them have more than 13 galaxies. Nine such groups lie in the HDC
of the supercluster, and 12 lie at the supercluster axis along an almost straight line.  
These groups in the supercluster, their galaxy content, and dynamical state
were analysed in \citet{2018A&A...620A.149E}. 
We highlight groups with at least ten galaxies
in the cocoon to compare the properties of groups in both environments. 
We give the data about groups in the  HDC in Table~\ref{tab:gr10} together
with data about groups in the cocoon, denoted as C1--C4.
The location of these groups in the sky plane is shown in Fig.~\ref{fig:radecvelfil}.

In Table~\ref{tab:gr10}, we also provide data on groups denoted as $G_{gap}$ and
 $G_{VO}$. %, and $G_{RQ}$. 
Group $G_{gap}$ has the largest magnitude gap between its two brightest
galaxies among cocoon groups, $|\Delta M_{12}| = 1.45$. 
This group is also 
interesting because of the environment; we discuss the group $G_{gap}$
in Sect.~\ref{sect:con}. 
The group $G_{VO}$ is an example of a poor group in which
six out of seven group member galaxies have very old stellar populations
(we define galaxy populations used in our analysis below 
 in Sect.~\ref{sect:skyvel}).

\begin{figure}[ht]
\centering
\resizebox{0.38\textwidth}{!}{\includegraphics[angle=0]{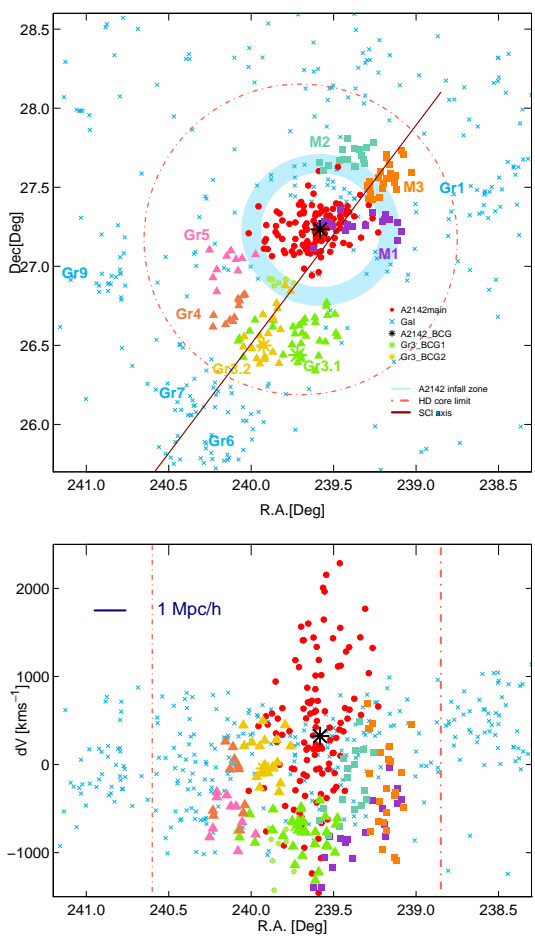}}
\caption{
Distribution of galaxies of the supercluster SCl~A2142 HDC in the sky plane
(upper panel),
and in the RA - velocity difference plane (lower panel).  
The red symbols show galaxies with old stellar populations
($D_n(4000) \geq 1.55$), and the blue symbols denote galaxies with young 
stellar populations ($D_n(4000) < 1.55$).
Red filled circles correspond to galaxies in the main component of the
cluster A2142. Blue crosses show other galaxies in the HDC. 
Filled squares show galaxies in infalling subclusters M1--M3
(violet - M1, pale green - M2, and orange - M3), and 
filled triangles denote galaxies in HDC groups Gr3--Gr5 
(Table~\ref{tab:gr10};
violet triangles - Gr5, orange triangles - Gr4, green and yellow triangles denote
two components of Gr3, as explained in the text).
Lines show HDC boundaries and the direction of the supercluster
axis, and the blue circle marks the infall zone of the cluster A2142.
}
\label{fig:hdcradec}
\end{figure}

\begin{figure}[ht]
\centering
\resizebox{0.47\textwidth}{!}{\includegraphics[angle=0]{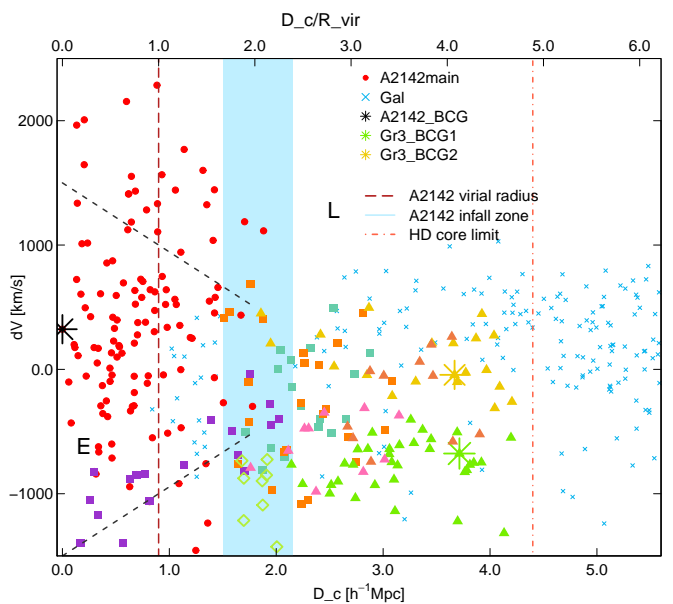}}
\caption{
Velocity of galaxies with respect to the cluster mean velocity vs. 
projected distance from the centre of the A2142 cluster ($D_c$)
in the highest density core of the supercluster up to 
clustercentric distances $D_c \leq 5.2$~\Mpc\ (PPS diagram,
this distance range covers the entire high density core 
of the supercluster, seen in Fig.~\ref{fig:hdcradec}).
Lines which separate early and late infall regions are
calculated using radius $1.8$~\Mpc\ \citep{2018A&A...610A..82E}. 
Notations are as in Fig.~\ref{fig:hdcradec}. 
Blue polygon marks infall zone of the cluster A2142,
dark red dashed line denotes the virial radius of the cluster,
and light red dot-dashed line denotes the boundaries of the HDC
of the supercluster.
}
\label{fig:hdcpps}
\end{figure}

To analyse the evolutionary state of groups, several methods can be used, 
substructure analysis among them. However, 
groups C1--C4 have less than 20 galaxies, and it is not straightforward
to analyse their substructure \citep[as shown, for example, by ][]{2013MNRAS.434..784R}. 
They lie away from each other
and do not form possibly merging group pairs, as do some groups in the supercluster
along the supercluster tail
\citep{2018A&A...620A.149E}. 
Thus, we use magnitude gap between group's two brightest
galaxies, $|\Delta M_{12}|$ as an indicator of their evolutionary state 
\citep{2010MNRAS.405.1873D, 2016A&A...586A..40K}. As found in these studies,
simulations show a trend that 
the magnitude gap $|\Delta M_{12}|$ is larger in groups which have been formed
earlier than groups with smaller magnitude gaps. In addition, they found that
the magnitude gap
between the brightest, and third or fourth brightest galaxy, $|\Delta M_{13}|$ and 
$|\Delta M_{14}|$ can also be used as an indicator of early formed groups. 
A good  example of such case from observations is the cluster A2142.
Several X-ray and optical studies have shown that in the cluster A2142
the second brightest galaxy is a member of recently infallen group,
and to characterise the cluster,  $|\Delta M_{13}|$ is a better indicator
\citep[see Table~\ref{tab:gr10} and discussion in][]{2000ApJ...541..542M, 2018A&A...610A..82E}.
In X-ray studies the magnitude gap  $|\Delta M_{12}|$ is used to define possible 
fossil groups as X-ray groups with $|\Delta M_{12}| \geq 2$ \citep{1994Natur.369..462P}.
Table~\ref{tab:gr10} shows that  the value of $|\Delta M_{12}|$ for groups is larger than
$1$ for one group.
There are no groups with $|\Delta M_{12}| \geq 2$ in our sample 
but for $G_{gap}$ $|\Delta M_{13}| = 2.25$. As already mentioned, we discuss this group in
Sect.~\ref{sect:con}.

Next we briefly present data on the structure of the  
{\bf HDC} of SCl~A2142. We use this information to determine its connectivity.
The HDC is seen in Fig.~\ref{fig:radecvelfil} as the innermost
region of the supercluster main body. 
HDC has radius of about $5$~\Mpc, centred at $R.A. =239.5$\degr and $Dec. = 27.3$\degr.
In the luminosity-density distribution of SCl~A2142, it is seen as the region of the highest
luminosity-density in the supercluster, with $D8 > 17$
\citep[][]{2015A&A...580A..69E, 2015A&A...581A.135G}.
HDC embeds eight galaxy groups with at least ten
member galaxies; we list them in Table~\ref{tab:gr10}. Here, we use
the same numbering of groups as in \citet{2018A&A...620A.149E}. 
In addition to groups which surround the cluster A2142 in the HDC, 
\citet{2018A&A...610A..82E} showed that the cluster A2142 
has three substructures, which they denoted as M1--M3
(where M comes from merging). In the group catalogue by \citet{2014A&A...566A...1T}
galaxies in M1--M3 are members of the cluster A2142. The analysis of the substructure and dynamical
state of the cluster revealed that these structures are probably the remnants
of infalling groups and filaments, which are already connected to the cluster
and therefore the cluster finding algorithm joined the galaxies in them with the
main cluster. For details of this analysis, we refer to \citet{2018A&A...610A..82E}.
The HDC is  elongated along the supercluster axis (Figs~\ref{fig:radecvel} and 
\ref{fig:hdcradec}).  

We plot the sky distribution of galaxies in different
structures of the HDC in the upper panel of Fig.~\ref{fig:hdcradec}. 
In the lower panel of Fig.~\ref{fig:hdcradec}, high velocities of galaxies in the
main cluster is a signature of the Finger-of-God effect in the cluster.
Galaxies in infalling groups (Gr 3, 4, and 5 in Table~\ref{tab:gr10})
are shown with different colours. The group 
Gr3 probably consists of two close
groups combined together by group finding algorithm.
In Fig.~\ref{fig:hdcradec}, these components are denoted with green and yellow triangles.
We also show galaxies in substructures M1--M3 in different colours.
In Figs.~\ref{fig:hdcradec} zone at which 
substructures fall into A2142 cluster is shown in blue
\citep[see][for details]{2018A&A...610A..82E}. The infall zone extends approximately
up to $2$~\Mpc\ from the cluster centre. 
This agrees with the clustercentric
distance limit for the cluster members in the caustic diagram by \citet{2018ApJ...863..102L}. 
In this region, the stellar ages and
star-formation rates of galaxies change rapidly \citep{2018A&A...610A..82E}. 
Moreover, simulations suggest that some
galaxies in the infall zone of galaxy clusters may actually be backsplash galaxies, meaning galaxies which entered the cluster a long time ago and  are now moving
out of the cluster \citep{2015ApJ...806..101H, 2017ApJ...843..128R}.  
The blue area with a radius approximately in the range of $1.5 - 2.2$~\Mpc\
in Fig. \ref{fig:hdcradec} is the most likely region for
backsplash galaxies in A2142.

To characterise the dynamical state of the HDC, we present its
projected phase space (PPS) diagram (Fig.~\ref{fig:hdcpps}).
In this figure, we mark the virial radius of the cluster with the dark red
dashed line, an infall zone of A2142 with the blue rectangle, and 
the border of the HDC with the light red dash-dotted line.
In PPS diagram  line-of-sight velocities of galaxies 
with respect  to the cluster mean velocity are plotted against 
the projected clustercentric distance (distance from the centre of A2142 in our case). 
Simulations show that in the PPS diagram 
galaxies at small clustercentric 
radii form an early infall (virialised) population with 
infall times $\tau_{\mathrm{inf}} > 1$~Gyr, 
and galaxies 
at large clustercentric radii 
form late or ongoing infall populations with $\tau_{\mathrm{inf}} < 1$~Gyr
\citep{2013MNRAS.431.2307O, 2014ApJ...796...65M, 2015ApJ...806..101H, 
2017ApJ...843..128R, 2019MNRAS.484.1702P, 2019ApJ...876..145S}. The early infall zone E in A2142 corresponds approximately to the region between
blue dashed lines in Fig.~\ref{fig:hdcpps}.
The late infall zone is indicated with L (from late).

Figure~\ref{fig:hdcpps} shows that member galaxies of 
substructures and groups near
the cluster A2142 populate the late infall zone of the PPS diagram 
up to clustercentric distances
$3-4$~\Mpc\ 
\citep{2018A&A...610A..82E, 2018A&A...620A.149E}.
This distance set as the distance limit for cluster members is seen also
in \citet{2013ApJ...764...58G}.
At the projected distance from the centre of the A2142 cluster of about
$D_c \approx 4.5$~\Mpc,\ the distribution of galaxies has a break
(Fig.~\ref{fig:hdcradec}  and ~\ref{fig:hdcpps}). This minimum 
corresponds to the luminosity-density limit $D8 = 17$ and outlines
the HDC of SCl~A2142.

\section{Connectivity in the supercluster and in the cocoon}
\label{sect:conn}  

The connectivity of a cluster or group is defined 
as the number of filaments connected to it.
Next we study the connectivity of the HDC of the superluster,
of the whole supercluster, and 
of groups with at least ten member galaxies in the cocoon. 
To find the connectivity of galaxy groups, we study filament membership of their
galaxies.

The cluster A2142 has six-seven structures connected to it. 
Figure~\ref{fig:hdcradec} shows that
infalling structures around the cluster A2142 extend up to clustercentric
radius of about $5$~\Mpc. 
At this location, there is a minimum in the luminosity-density distribution 
in the supercluster \citep[this was shown in Fig. 5 in ][]{2015A&A...580A..69E}.
We can suppose that the HDC of the supercluster
is detached from the surrounding supercluster core due to the dynamical 
processes during collapse which also might have destroyed 
filament possibly associated with Gr9 (see below).

Figure~\ref{fig:radecvelfil} shows that filaments in the outskirts region of 
the supercluster main body and around it are short, 
and they do not reach the  HDC of the supercluster.
Six long filaments with length over $20$~\Mpc\ begin near the supercluster
main body in the cocoon but they do not have common member galaxies with the 
supercluster. 
One long filament lies at the extension of the
supercluster tail. Therefore, the connectivity of the supercluster main body 
is $\pazocal{C} = 6$, and for the whole supercluster, it is $\pazocal{C} = 7$. 

Table~\ref{tab:gr10} shows that groups in the cocoon which lie on filaments have 
a connectivity $\pazocal{C} = 1$.
In the HDC two groups (Gr3 and Gr4), lie on the supercluster axis.
Groups Gr5 and Gr9 may belong to a small filaments, therefore,
we consider their connectivity to also be $\pazocal{C} = 1$
\citep[see][for details]{2018A&A...620A.149E}. 

It is interesting to compare these results with other studies
of the connectivity of groups. However, a word of caution is needed at the start.
If a group lies in a filament, then we considered in this study
that its connectivity is $\pazocal{C} = 1$, if filaments going
out of the group from different sides belong to the same filament
in the filament catalogue. We consider $\pazocal{C} = 2$, if they are 
different filaments. In some studies 
$\pazocal{C} = 2$ in both cases  \citep{2018MNRAS.479..973C}. 
Also, as discussed above, different filament finding methods and
criteria for filament membership may give different results about
the filament membership.

Altogether, this means  that 
exact comparison of group connectivities from different studies
may not be straightforward.
Thus, we only note that we found trends in group connectivities
similar to those found, for example, by \citet{2019MNRAS.489.5695D}
for poor groups at high redshifts.
For low-mass groups from COSMOS data at high redshifts 
\citet{2019MNRAS.489.5695D} found that, on average,
their connectivity $\pazocal{C} \approx 2$,
higher than the connectivity of groups in SCl~A2142 cocoon.
They showed that the connectivity of 
most massive  X-Ray detected groups in COSMOS field is, on 
average, up to 4, which is slightly lower than for A2142.
The difference may be related to group mass (and richness), as
the connectivity is higher for high mass (and richer) groups
\citep{2019MNRAS.489.5695D, 2020A&A...635A.195G}.

\section{Galaxy populations in the supercluster cocoon}
\label{sect:galpop}

Now we analyse the spatial distribution of galaxies with different star 
formation histories and focus on blue star forming (BSF) galaxies,
on recently quenched (RQ) galaxies,
on  red star forming (RSF) galaxies, and on galaxies with very old stellar populations
(VO). These populations are obtained using galaxy data which were 
described in Sect.~\ref{sect:galpop}. 
To explain how studied  galaxy populations are defined,
we show in Fig.~\ref{fig:dn}
the distributions of  $D_n(4000)$ index values 
and colour index $(g - r)_0$ values for galaxies
in the supercluster and in the cocoon, and separately for galaxies in groups and for single 
galaxies. Figure~\ref{fig:dn} also provides 
the first comparison between galaxy content of the supercluster and the cocoon,

The left panel of Fig.~\ref{fig:dn} shows that the distribution of 
the $D_n(4000)$ index values have a minimum at
$D_n(4000) \approx 1.55$. 
\citet{2003MNRAS.341...33K} showed that the value $D_n(4000) = 1.55$ corresponds to 
the mean age of about $1.5$~Gyr.
As in \citet{2003MNRAS.341...54K, 2017A&A...605A...4H, 2018A&A...620A.149E},
we use this limit to separate galaxies with old and young
stellar populations.
Galaxies with young stellar populations have  $D_n(4000) \leq 1.55$.

VO galaxies  are defined as having  $D_n(4000) \geq 1.75$.
At this value, there is a drop in the star-formation rate of galaxies 
\citep[see Fig. 27 in ][]{2004MNRAS.351.1151B}. 
According to \citet{2003MNRAS.341...33K} the value $D_n(4000) = 1.75$ corresponds to 
the mean age of about $4$~Gyr (for Solar metallicity) or older 
(for lower metallicities). The centre of the cluster A2142 
is populated by such galaxies \citep{2018A&A...610A..82E}.
We could expect that such galaxies are typical for very dense environments.
We searched for VO galaxies 
in the supercluster cocoon, with the expectation that
such galaxies are (nearly) absent in poor groups in the low global density environments. 
Six galaxies with $D_n(4000) \geq 1.75$ have a star-formation rate $\log \mathrm{SFR} > -0.5$;
they were excluded from this sample
to maintain  a sample of galaxies without ongoing star formation.

In the right panel of Fig.~\ref{fig:dn}, red and blue galaxies are
approximately separated by the colour index value $(g - r)_0 = 0.7$; red galaxies 
have $(g - r)_0 \geq 0.7$. 
This limit depends on the luminosity of galaxies, but since we do not have very faint 
galaxies in our sample, we may use such a simple approach.

The BSF, RSF, and RQ galaxies are defined using a combination of parameters.
The BSF galaxies are defined as having blue colours 
($(g - r)_0 < 0.7$)  and a star formation rate $\log \mathrm{SFR} \geq -0.5$. 
The RSF galaxies are defined  as red galaxies with colour index
$(g - r)_0 \geq 0.7$,  and star formation rate $\log \mathrm{SFR} \geq -0.5$.
The RQ galaxies are defined as galaxies with low SFRs
and low values of $D_n(4000)$ index: 
$D_n(4000) \leq 1.55$ and  $\log \mathrm{SFR} < -0.5$. 
For details about these limits, we refer to 
\citet{2018A&A...610A..82E, 2018A&A...620A.149E}. 
Data on galaxy populations are summarised in Table~\ref{tab:galpopdef}.

\begin{table*}[ht]
\caption{Data on galaxy populations.}
\begin{tabular}{lllr} 
\hline\hline  
(1)&(2)&(3)&(4)\\      
\hline 
Population & Abbr. &Definition&  $N_{\mathrm{gal}}$ \\
\hline                                                    
Blue star forming galaxies  & BSF &  $(g - r)_0 < 0.7$, $\log \mathrm{SFR} \geq -0.5$      & 295  \\
Red star forming galaxies   & RSF &  $(g - r)_0 \geq 0.7$, $\log \mathrm{SFR} \geq -0.5$   & 184  \\
Recently quenched galaxies  & RQ  &  $D_n(4000) \leq 1.55$,  $\log \mathrm{SFR} < -0.5$    & 30  \\
Galaxies with very old stellar populations & VO &  $D_n(4000) \geq 1.75$ & 331  \\
\hline
\label{tab:galpopdef}  
\end{tabular}\\
\tablefoot{                                                                                 
Columns are as follows:
(1): Galaxy population;
(2): Abbreviation;
(3): Definition of a given population;
(4): The number of galaxies in a given population in the cocoon. 
}
\end{table*}

\begin{figure*}[ht]
\centering
\resizebox{0.42\textwidth}{!}{\includegraphics[angle=0]{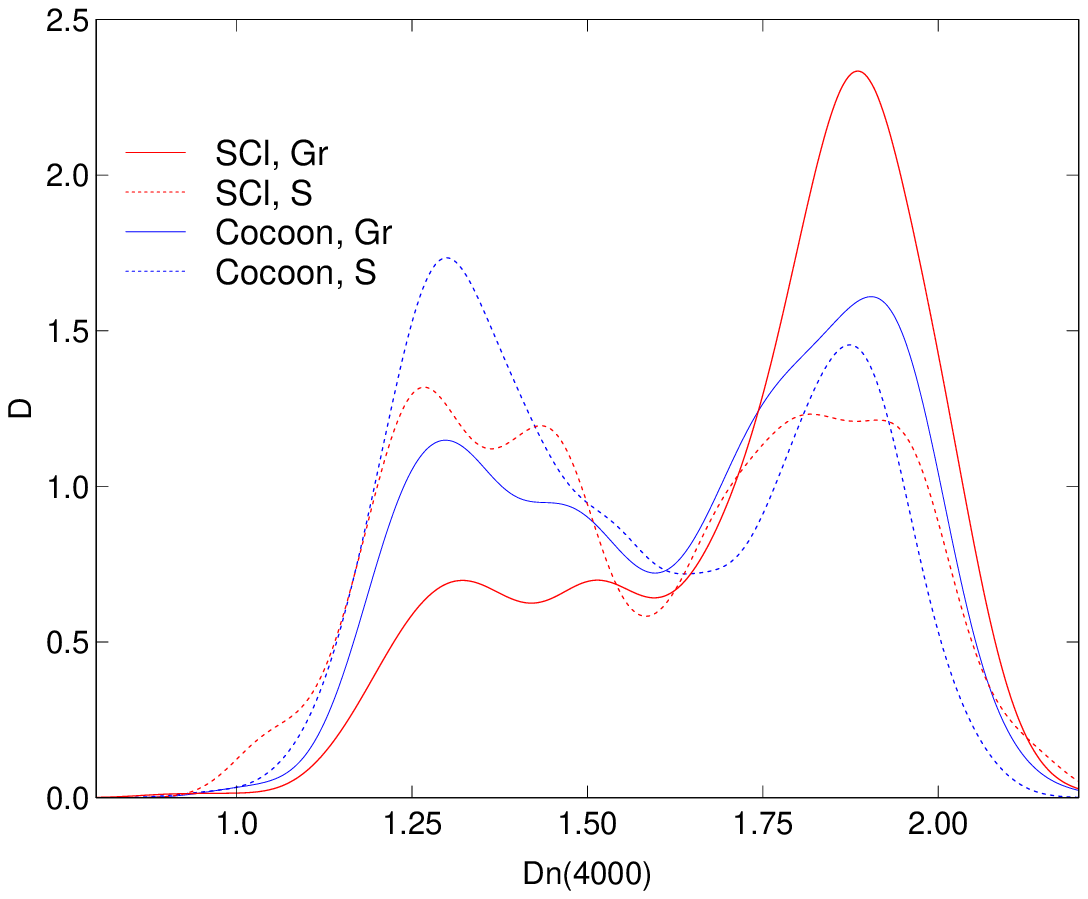}}
\resizebox{0.42\textwidth}{!}{\includegraphics[angle=0]{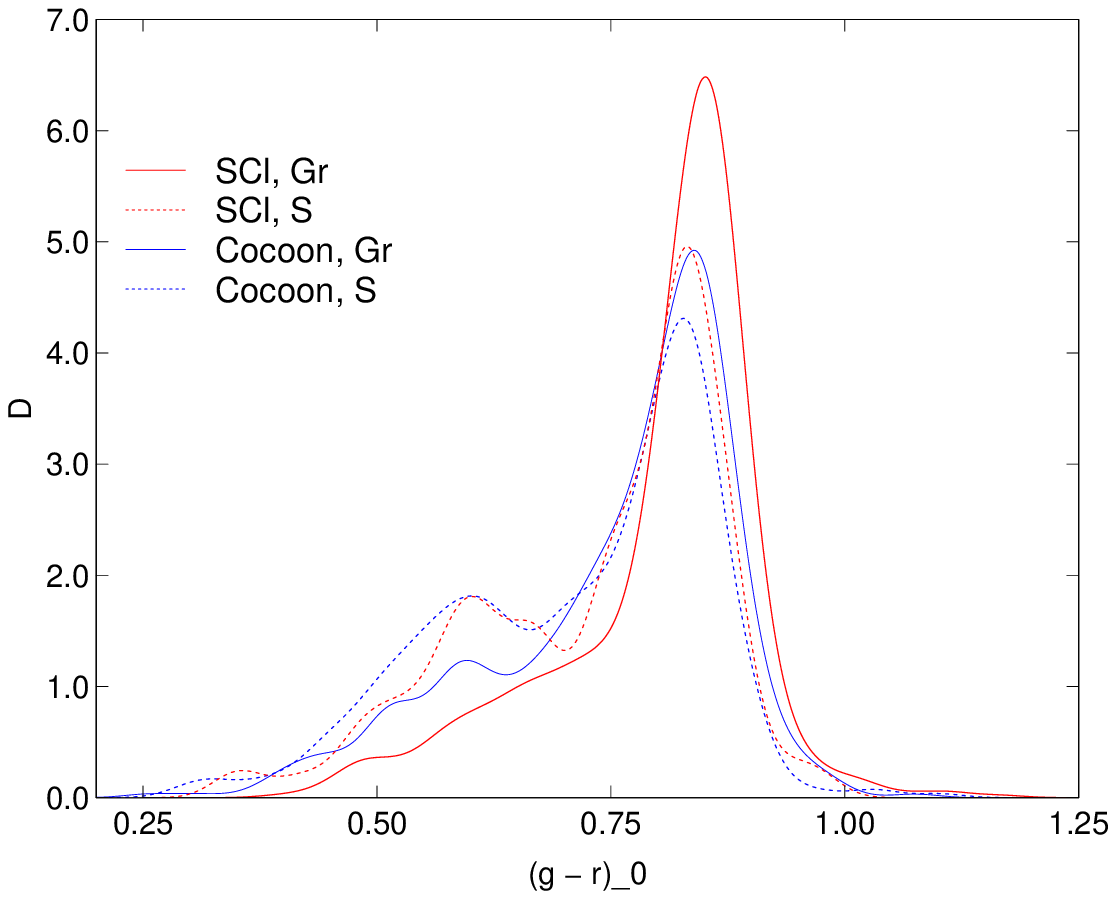}}
\caption{
Distributions of  $D_n(4000)$ index values (left panel)
and colour index values $(g - r)_0$ (right panel) for galaxies in groups (Gr, solid lines)
and for single galaxies (S, dashed lines)
in the supercluster (red lines) and in the cocoon (blue lines).
}
\label{fig:dn}
\end{figure*}

Next we analyse the distribution of galaxies from these populations. 
At first, we compare overall galaxy content of the supercluster and cocoon.
Figure~\ref{fig:dn} shows that galaxy groups in the supercluster contain relatively more
galaxies with high values of $D_n(4000)$ index than those in low  global 
density region around
it, and also relatively more red galaxies with $(g - r)_0 \geq 0.7$. 
Groups in the supercluster contain relatively more passive galaxies
than groups in the cocoon (74\% and 60\%, accordingly).
In the supercluster, VO galaxies make up 60\% of galaxies in groups, and in the cocoon, 40\% of group members. These galaxies
form 1/3 of single galaxy population in both environments.

The Kolmogorov-Smirnov
(KS) test showed that differences in galaxy content of groups
between supercluster and cocoon groups are significant at very high 
level, with $p$-values (the estimated probability of rejecting the hypothesis
that distributions are statistically similar) $p < 0.001$. 
The difference between $D_n(4000)$ index values 
and colour index $(g - r)_0$ values for single galaxies in both environments
are not statistically significant.
This shows the importance of the group environment in which 
galaxies are transformed from BSF galaxies to VO galaxies. Figure~\ref{fig:dn}
suggests that transformations of galaxies are more efficient in high-global-density
environment.

Figure~\ref{fig:dn} also  shows that $D_n(4000)$ index is a much more
sensitive proxy for the age of the stellar populations in galaxies
than broad band colours. This is because 
the definition of the $D_n(4000)$ index used in this paper
is designed to be much less sensitive to reddening. 
Extinction by dust, that is, reddening, has the effect of 
reducing the broad-band colour variation since the 
star-forming galaxies, which are naturally bluer, 
suffer stronger extinction, which reddens their integral light.

\subsection{Distribution and local density of galaxies from various populations}
\label{sect:skyvel} 

We show the  sky distribution 
of galaxies from populations being studied in the cocoon 
in Fig.~\ref{fig:radecpop}. 
Star-forming galaxies (both BSF and RSF) in Fig.~\ref{fig:radecpop} 
follow evenly the structures in the cocoon.
Most RQ galaxies
lie at low declinations  with $Dec < 20$~degrees, and
$240 < R.A. < 246$,
and they are absent elsewhere. 
The VO galaxies in Fig.~\ref{fig:radecpop} can be  seen almost 
in all structures of the low-global-density environment, even among single galaxies.
In some very poor groups in the cocoon almost all galaxies
have very old stellar populations. 
Group denoted as $G_{VO}$ in Table~\ref{tab:gr10} is an 
example of a poor group in which six (out of seven) member galaxies
are VO galaxies.

\begin{figure}[ht]
\centering
\resizebox{0.46\textwidth}{!}{\includegraphics[angle=0]{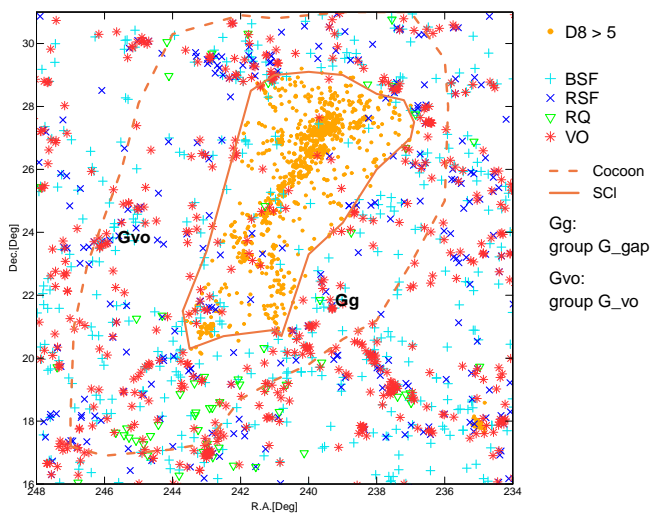}}
\caption{
Distribution of BSF (turquoise crosses), RSF (blue x-s), RQ
(green triangles), and VO (dark red stars) galaxies in the SCl~A2142 
cocoon in the sky plane. Notations
are the same as in Fig.~\ref{fig:radecvel}. 
Orange dashed line shows the cocoon boundaries, and orange solid line shows
the supercluster boundaries.
}
\label{fig:radecpop}
\end{figure}

\begin{figure*}[ht]
\centering
\resizebox{0.42\textwidth}{!}{\includegraphics[angle=0]{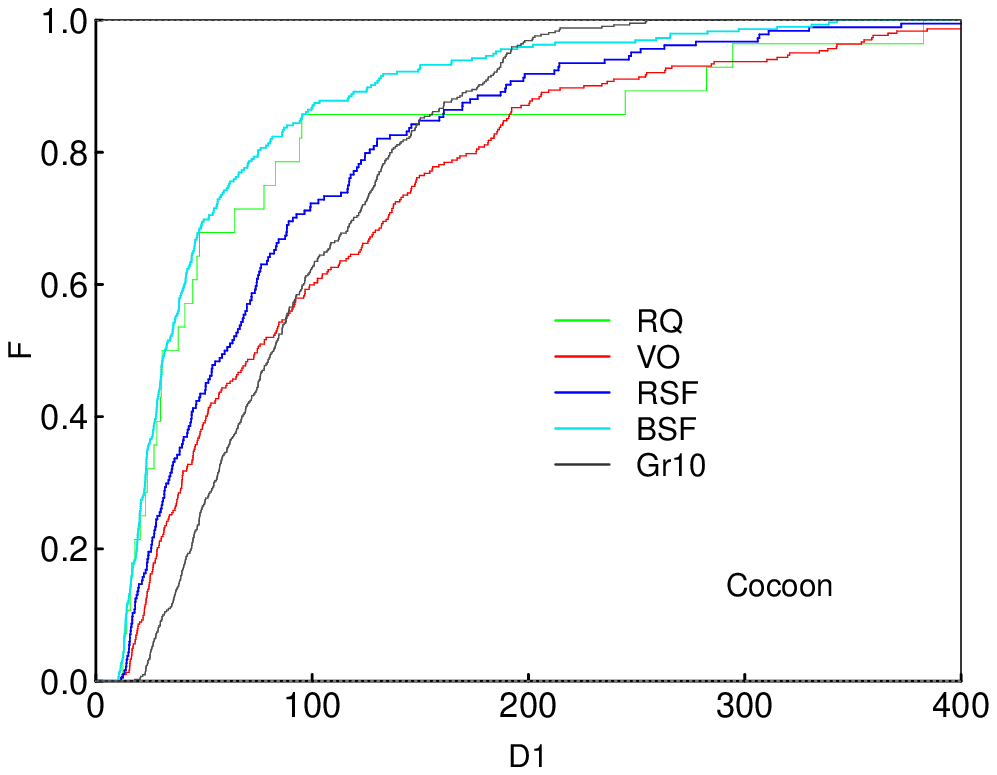}}
\resizebox{0.42\textwidth}{!}{\includegraphics[angle=0]{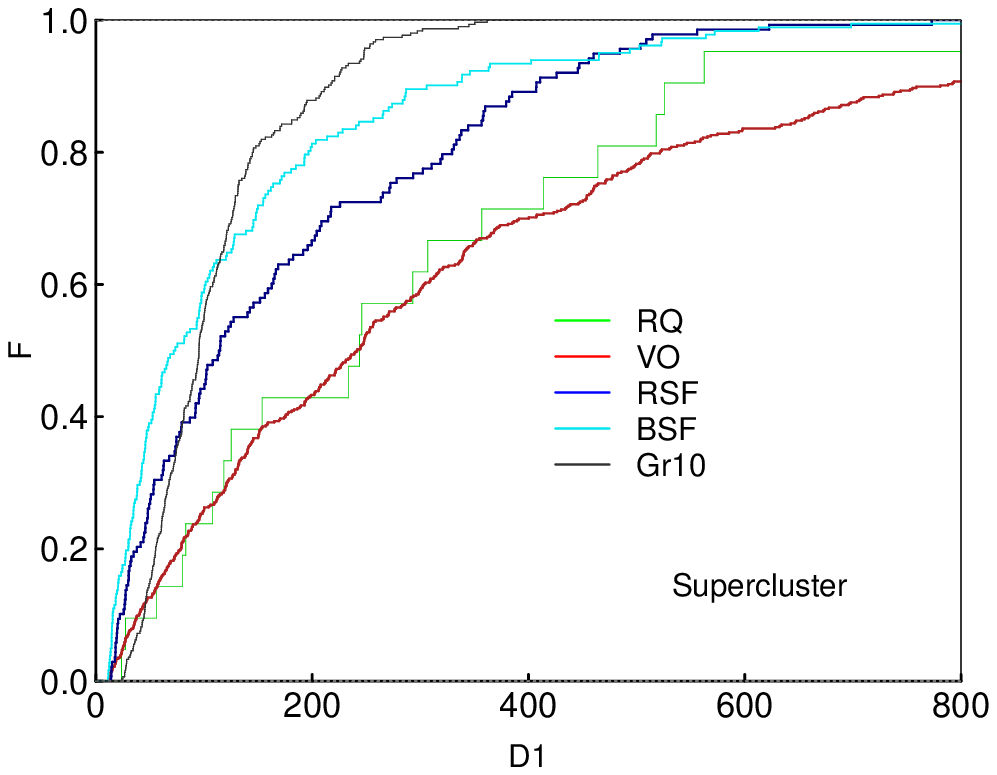}}
\caption{
Cumulative distributions of the local luminosity-density field values $D1$ around BSF galaxies (turquoise line),
RSF galaxies (blue line),  RQ galaxies (green line), and VO galaxies (red line)
in the supercluster (right panel) and in the cocoon (left panel).
Grey line shows  $D1$ distributions for poor groups with $N_{gal} \leq 10$.
Note the different $D1$ range in the cocoon (left panel, $0 \leq D1 \leq 400$),
and in the supercluster (right panel, $D1$ spans
from $0$ to $800$). 
}
\label{fig:d1sclc}
\end{figure*}

To quantify the differences in the distribution of galaxies from
populations in this study, we calculated the local luminosity-density values around galaxies, $D1.$ 
The distributions of $D1$ are
presented in Fig.~\ref{fig:d1sclc} (in the left panel for galaxies in the cocoon and 
in the right panel for supercluster galaxies). 
The median values of the local luminosity-density around BSF, RSF, RQ, and VO galaxies 
(in mean luminosity-density units; see Sect.~\ref{sect:cocoon}), along with the $p$-value from the KS test, in which we compared $D1$ from a 
given populations in the supercluster and in the cocoon,
are given in Table~\ref{tab:galpopc}.
As seen from Fig.~\ref{fig:d1sclc} and Table~\ref{tab:galpopc},
in the supercluster  
median D1 values of BSF, RSF, RQ, and VO galaxies  follow 
the respective sequence of star-formation activity -- 
in denser environments, the current star-formation activity is lower. 
However, the situation is somewhat different in the cocoon. 
In the cocoon, on average, RSF galaxies lie in denser environments 
than do RQ galaxies ($60$ versus $34$). Thus, star-formation quenching in 
RQ galaxies in the cocoon may exhibit a different mechanism 
when compared with the supercluster.
We return to this issue in Sect.~\ref{sect:discussion}.

Low values of local densities $D1$
at locations of BSF and $RQ$ galaxies in the cocoon  suggest that both populations 
lie in outer parts of systems (groups and filaments). 
In the supercluster, local densities for BSF galaxies
are the lowest among galaxy populations in the supercluster.
\citet{2018A&A...620A.149E} found that in SCl~A2142, star-forming galaxies
lie on the outskirts of groups and in poor groups (or
they are single galaxies) in the outskirts of the supercluster
main body, which explains their low local densities. 
RSF galaxies lie at intermediate local densities, often
in infall regions of groups and clusters, or along filaments
\citep{2009MNRAS.399..966S, 2018A&A...620A.149E}.
However, in the supercluster, local densities $D1$ for $RQ$ galaxies are
almost as high as for VO galaxies, which is in agreement with findings in 
\citet{2018A&A...620A.149E} stating that these galaxies lie on the outskirts of
groups  and of the cluster A2142 where local densities are high.
Some of them lie also in the outskirts of the supercluster
main body.

The local
density around VO galaxies in the cocoon is the highest among cocoon populations.
They occupy inner parts of groups and filaments. 
High local densities at the location of VO galaxies in the supercluster
come from the fact that 
these galaxies lie in the central part of the cluster A2142 or in the centres of
other rich groups,
as well as from the higher local densities around poor groups in the supercluster
in comparison with those in the cocoon (Fig.~\ref{fig:d1sclc}).

\subsection{Properties of galaxies from various populations}
\label{sect:galprop}

Next we compared other properties of galaxies from these populations,
namely the stellar masses, ${\mathrm{log}} M^{\mathrm{*}}$, concentration indexes, $C$, 
the probabilities of late type, $P_{\mathrm{late}}$, 
stellar velocity dispersions, $\sigma^{\mathrm{*}}$, and metallicities, $Z$.
The corresponding 
distributions are shown in Figs.~\ref{fig:rqvorssm} and \ref{fig:rqvorsp59}. 
Their median values and the $p$-value of the KS test are given in
Table~\ref{tab:galpopc}.

The distribution of stellar masses of galaxies in
Fig.~\ref{fig:rqvorssm} shows that BSF galaxies 
both in the cocoon and in the supercluster have significantly
lower stellar masses than galaxies from other populations. This is 
what could be expected from galaxies which are
still forming their stellar content and increasing their stellar mass.
Small differences in stellar masses of galaxies from other populations 
are not statistically significant
and we can conclude that galaxies from these populations in our
sample have statistically similar stellar masses, in a range of 
$M^{\mathrm{*}} \approx 3\times10^{9}h^{-1}M_\odot - 3\times10^{11}h^{-1}M_\odot$.

Figure~\ref{fig:rqvorsp59} (upper left panel) shows that BSF galaxies have the highest
values of  the concentration index, $C$,  
and VO galaxies have the lowest $C$ values among our galaxy populations,
as they are similar in the supercluster and in the cocoon. 
In this respect, RQ 
galaxies differ from others. In the supercluster they
have lower values of $C$  than in the cocoon, 
these differences  
are statistically significant at a very high level, with a $p$-value of
$p \leq 0.01$ (Table~\ref{tab:galpopc}). 

The distribution of probabilities for galaxies 
to be of late-type, $P_{\mathrm{late}}$, are plotted in Fig.~\ref{fig:rqvorsp59} (upper right panel).
This probability is the highest
for BSF galaxies and for RQ galaxies in the cocoon, and the lowest
for VO galaxies (in both environments).
However, KS test shows that these  differences are not statistically 
significant (Table~\ref{tab:galpopc}).

BSF galaxies also have the lowest
values of stellar velocity dispersions and VO galaxies have the highest 
(Fig.~\ref{fig:rqvorsp59}, lower left panel).
This agrees with earlier results which have shown that galaxies with young stellar
populations have lower stellar velocity dispersions than galaxies with old stellar
populations \citep{2009ApJ...694..867S, 
2012ApJ...760...62B, 2018A&A...610A..82E}.
The stellar  velocity dispersions
of RQ galaxies are similar in the supercluster and in the cocoon,
indicating, in accordance with the results for
$P_{\mathrm{late}}$, that the overall morphological type of RQ and RSF 
galaxies is the same in both environments.
As the stellar mass distribution for VO, RQ, and RSF samples are similar, 
this suggests that VO galaxies are mainly non-rotating 
early-type galaxies, but RQ and RSF galaxies (and BSF galaxies) 
may have also a significant rotation component corresponding 
to the dynamics of disky galaxies. 

The metallicities $Z$ of BSF, RQ, and RSF galaxies (Fig.~\ref{fig:rqvorsp59}, lower right panel)
are statistically similar in both environments,
BSF and RQ galaxies having the lowest metallicity values. VO galaxies in the supercluster
have higher
metallicities than galaxies from other populations in our study.
In the cocoon, they are similar to
the metallicities for galaxies from other populations. 
 
\begin{figure}[ht]
\centering
\resizebox{0.40\textwidth}{!}{\includegraphics[angle=0]{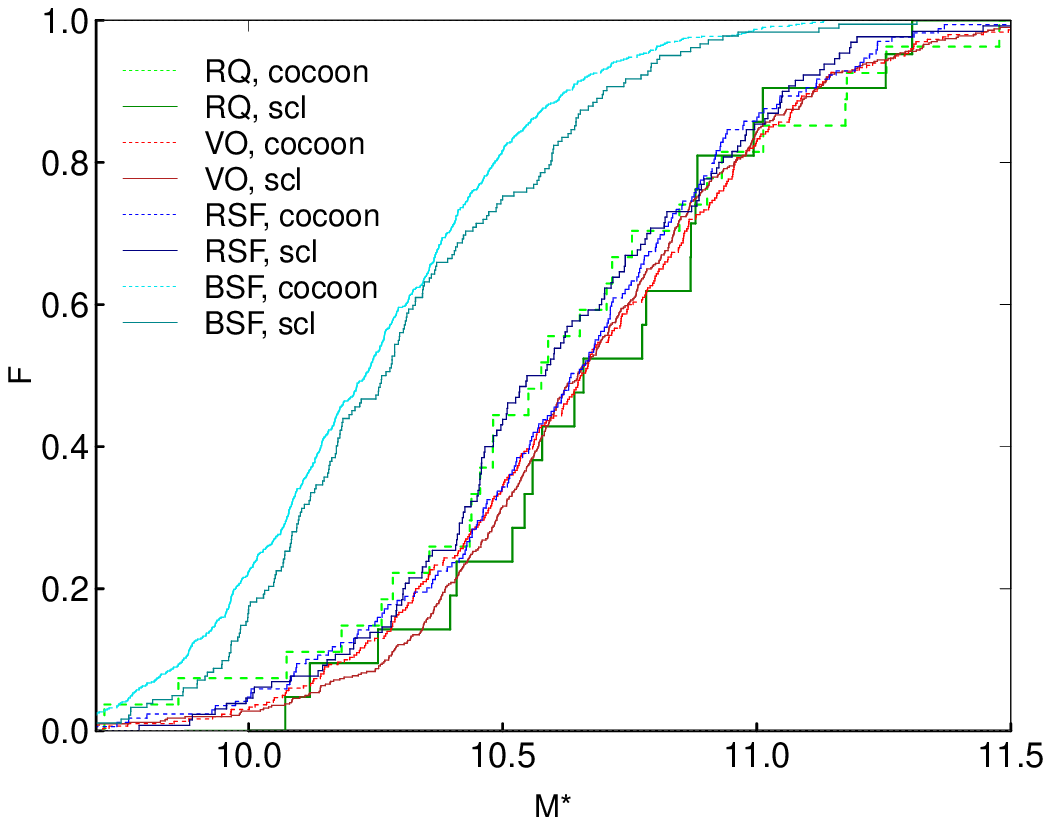}}
\caption{
Cumulative distribution of stellar masses ${\mathrm{log}} M^{\mathrm{*}}$ for 
BSF galaxies in the supercluster (solid dark turquoise line) and in the
cocoon (dashed light turquoise line), and the same for
RQ galaxies (solid dark green line and dashed light green line), for
RSF galaxies (dark solid and light dashed blue lines), and for
VO galaxies (dark solid and light dashed red lines).
}
\label{fig:rqvorssm}
\end{figure}

\begin{figure}[ht]
\centering
\resizebox{0.23\textwidth}{!}{\includegraphics[angle=0]{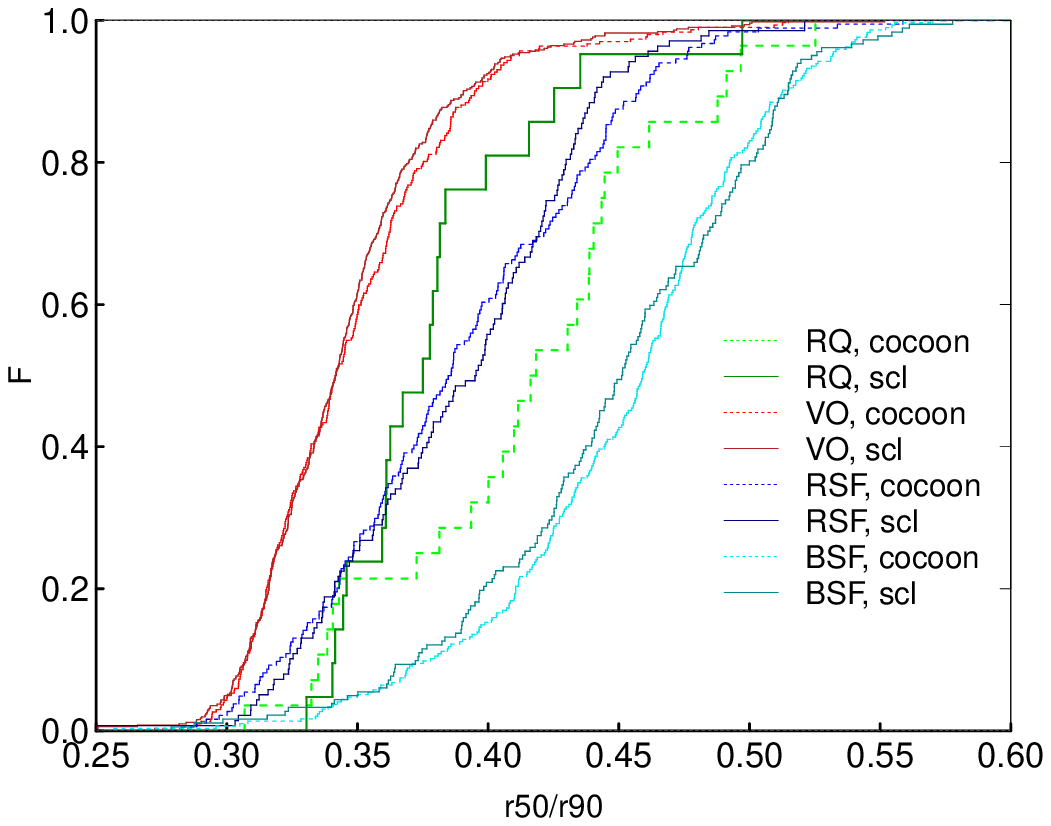}}
\resizebox{0.23\textwidth}{!}{\includegraphics[angle=0]{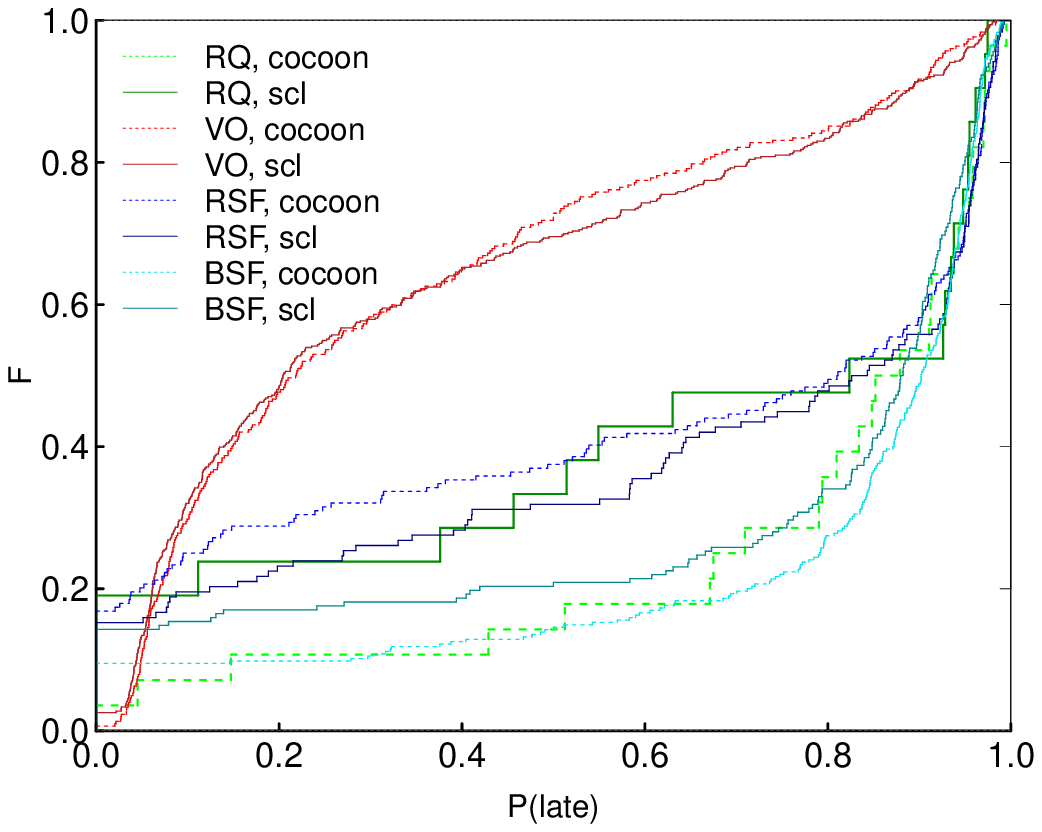}}\\
\resizebox{0.23\textwidth}{!}{\includegraphics[angle=0]{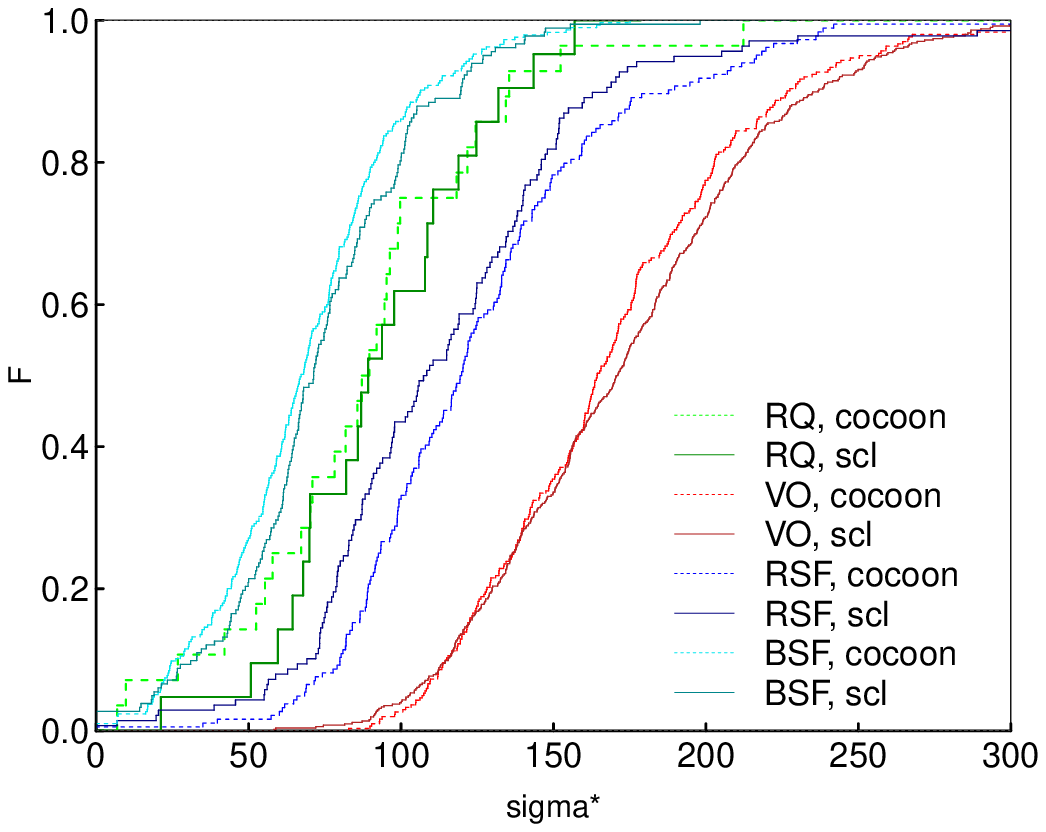}}
\resizebox{0.23\textwidth}{!}{\includegraphics[angle=0]{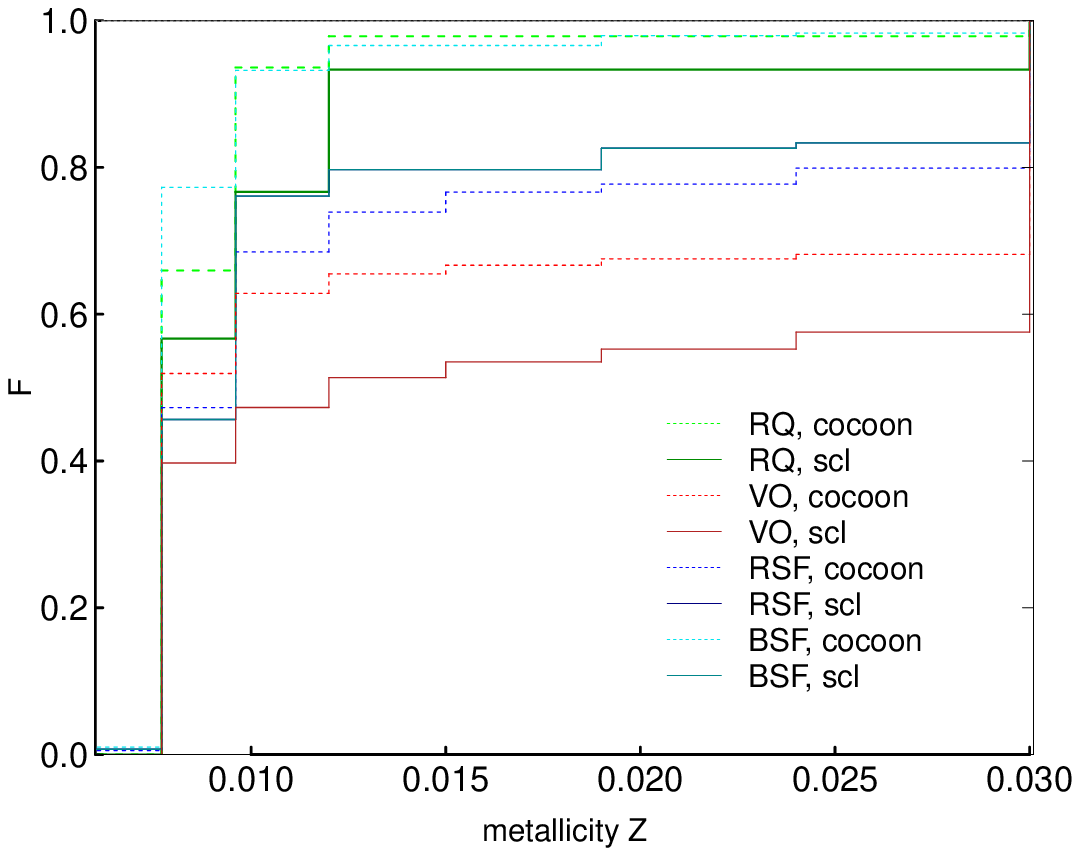}}
\caption{
Cumulative distribution of concentration indices,  $C = R_{50}/R_{90}$, probabilities to be of
late type, $P_{\mathrm{late}}$, 
stellar velocity dispersions, $\sigma^{\mathrm{*}}$ (in ${\mathrm{km s^{-1}}}$), 
and metallicities, $Z,$ for 
BSF galaxies in the supercluster (solid dark turquoise line) and in the
cocoon (dashed light turquoise line), and the same for
RQ galaxies (solid dark green line and dashed light green line), for
RSF galaxies (dark solid and light dashed blue lines), and for
VO galaxies (dark solid and light dashed red lines).
}
\label{fig:rqvorsp59}
\end{figure}

\begin{table*}[ht]
\caption{Median values of the local luminosity-density $D1$
around galaxies, and of galaxy properties,  and the KS test $p$-values
for RQ, RSF, BSF, and VO galaxies in the supercluster
(scl) and in the cocoon (c)}
\begin{tabular}{lrrrrrrrrrrr} 
\hline\hline  
(1)&(2)&(3)&(4)&(5)&(6)&(7)&(8)&(9)&(10)&(11)&(12)\\      
\hline 
 $Pop$ & $N_{\mathrm{gal}}$ & $D1$ & $p$ &  $C_{\mathrm{med}}$ & $p$ & $\sigma^{\mathrm{*}}_{\mathrm{med}}$ & $p$& 
  $P_{\mathrm{late}}^{\mathrm{med}}$ & $p$ & $Z_{\mathrm{med}}$ & $p$ \\
\hline
%                                                sigma*               hcs     met
 $BSF_{scl}$   & 182 & 73  & $< 0.001$&0.46  & 0.28 &  71  & 0.28&  0.88  & 0.05 &  0.0077  & 1.0  \\
 $BSF_{c}$   & 295 &  32 &          &0.46  &      &  67  &     &  0.90  &      &  0.0077  &       \\
 $RSF_{scl}$ & 138 & 114 & $< 0.001$&0.40  & 0.48 & 108  & 0.03&  0.83  & 0.51 &  0.0096  & 0.76  \\
 $RSF_{c}$   & 184 &  60 &          &0.39  &      & 120  &     &  0.81  &      &  0.0096  &       \\
 $RQ_{scl}$   &  47 & 244  & $< 0.001$&0.38  & 0.004& 89   & 0.79&  0.82  & 0.23 &  0.0077  & 0.67 \\
 $RQ_{c}$   &  30 &  34  &          &0.42  &      & 88   &     &  0.88  &      &  0.0077  &      \\
 $VO_{scl}$   & 516 & 242  & $< 0.001$&0.34  & 0.41 & 171  & 0.09&  0.21  & 0.69 &  0.0077  & $< 0.001$  \\
 $VO_{c}$   & 331 &  75  &          &0.34  &      & 164  &     &  0.20  &      &  0.0120  &       \\
\hline
\label{tab:galpopc}  
\end{tabular}\\
\tablefoot{                                                                                 
Columns are as follows:
(1): Population;
(2): Number of galaxies in a population;
(3--4): Median value of the local luminosity-density around a galaxy $D1$ and 
$p$-value of the KS test between a given population in
the supercluster (scl) and the cocoon (c);
(5--6): Median value of the concentration index $C$ and 
$p$-value of the KS test between a given population in
the supercluster and the cocoon; 
(7--8): Median value of the stellar velocity dispersion $\sigma^{\mathrm{*}}$ and 
$p$-value of the KS test between a given population in
the supercluster and the cocoon; 
(9--10): Median value of the probability to be of late type $P_{\mathrm{late}}$
\citep[sum of probabilities $P_{\mathrm{Sab}} + P_{\mathrm{Scd}}$
from][]{2011A&A...525A.157H}, and 
$p$-value of the KS test between a given population in
the supercluster  and the cocoon; 
(11--12): Median value of the metallicity $Z$ and 
$p$-value of the KS test between a given population in
the supercluster and the cocoon. 
}
\end{table*}

We additionally checked whether  the galaxy content of
long filaments with length $ \geq 20$~\Mpc\ depends on the distance 
of filament member galaxies from the supercluster centre, $D_c$.
The values of $D_n(4000)$ index versus $D_c$ for galaxies in long
filaments are plotted in Fig.~\ref{fig:fil20dn4}.
In this figure, background colours show the density of points
at corresponding  values of $D_n(4000)$ index and $D_c$.
The density of points (and therefore the number of galaxies) is the 
largest at clustercentric distances $D_c \approx 28$~\Mpc. 
Galaxies from these distances
lie in groups outside cocoon borders. 

We can see in Fig.~\ref{fig:fil20dn4} that galaxies in long filaments, which are the closest to 
the supercluster centre, lie at $D_c \approx 7$~\Mpc.  
In other words, long filaments do not reach the HDC of the supercluster, 
as we show in Sect.~\ref{sect:conn}.
Galaxies which are the closest to the supercluster centre are all
with old stellar populations. The nearest galaxies with young stellar
populations in long filaments lie on the outskirts 
of the supercluster at $D_c \approx 10$~\Mpc,
marked in the figure.
Distances $D_c \approx 20$~\Mpc\ correspond to
the cocoon borders around the supercluster main body, at larger distances,
galaxies belong to groups on long filaments outside the SCl~A2142 cocoon,
except those which lie in the straight filament at the extension
of the supercluster tail (filled red dots in the figure).
Figure~\ref{fig:fil20dn4} shows that 
there are no trends in star-formation properties along filaments
which cannot be explained by the changes in global or local density 
(membership of groups)
seen in Fig.~\ref{fig:dn}. 
This is similar to what \citet{2018A&A...620A.149E}
found for SCl~A2142. In the supercluster, galaxy content of groups which lie 
on the supercluster axis is statistically similar in the HDC and in the tail of the
supercluster, and it does not change significantly with the distance from the 
supercluster centre.

\begin{figure}[ht]
\centering
\resizebox{0.47\textwidth}{!}{\includegraphics[angle=0]{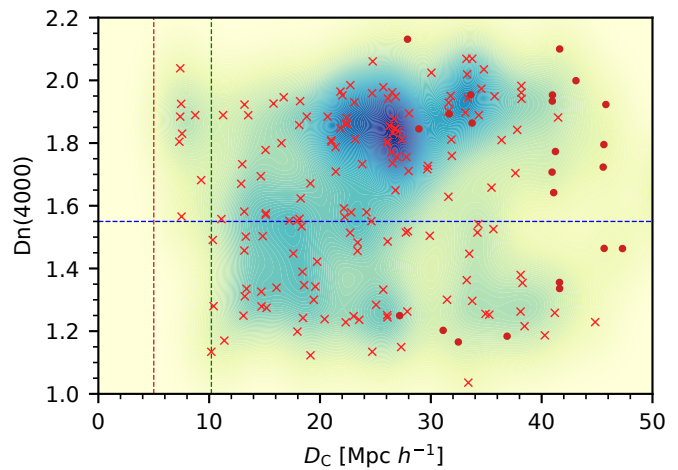}}
\caption{
$D_n(4000)$ index versus distance from the supercluster center ($D_c$) 
 for galaxies in
long filaments with length  $ \geq 20$~\Mpc\
(red crosses).
Filled red dots denote galaxies in the straight filament at the extension
of the supercluster axis (F1033).
Colours code the density of points at a given location in the plot.
We note that filaments around the supercluster main body begin at
clustercentric distances $D_c \approx 7$~\Mpc, and F1033 begins in the supercluster
tail at $D_c \approx 28$~\Mpc. 
Line at $D_n(4000) = 1.55$ separate galaxies with young ($D_n(4000) \leq 1.55$)
and old stellar populations. The line at $D_c = 10.2$~\Mpc\
show the distance from the supercluster centre for 
galaxies with young stellar 
populations  in long filaments which are the closest to the supercluster centre.
The line at $D_c = 5$~\Mpc\ shows the borders of the HDC of the supercluster. 
}
\label{fig:fil20dn4}
\end{figure}

\section{Discussion}
\label{sect:discussion}

Together with our earlier studies of the supercluster SCl~A2142
we find that 
the nearly spherical main body of the 
SCl~A2142 has an extension in the form of a quite long, 
about $30$~\Mpc\ straight tail. The supercluster tail 
forms an axis of the supercluster along which groups merge and
fall into the main cluster \citep[][and references therein]{2018A&A...620A.149E}. 
Also, cluster X-ray and radio haloes are elongated along
the supercluster axis. 
Thereafter, we found that this tail is further extended by a long 
filament of galaxies oriented nearly precisely along the axis of the supercluster. 
With a total length of $\approx 75$~\Mpc,\ they form the longest 
straight structure in the Universe found so far. 
Global luminosity-densities, $D8,$ change systematically along the structures 
in and around the supercluster 
from  
$D8 > 20$ within the HDC of the supercluster to $D8 < 2$ near the cocoon edge, and 
local luminosity-densities, $D1,$ span in the range of 
$D1 > 800$ within the HDC to $D1 < 10$ around the endpoint of the filament. 
Such a large density contrast between different ends of this structure makes it 
very useful to compare galaxy properties along it. 

In our study, we find (somewhat surprisingly) 
that while, in general, 
there are relatively more star-forming galaxies in the cocoon in
comparison with the supercluster, there are no strong global trends
in galaxy populations (BSF, RSF, RQ, and VO, in order of decreasing of
their current star-formation activity) 
along the straight structure with respect to supercluster centre. 
Instead, we find various environmental trends in the
star-formation properties of galaxies, which show that these properties depend
both on the local luminosity-density, $D1,$ and global luminosity-density, $D8$.
We discuss them below.

\subsection{Galaxy evolution in the supercluster
and in the cocoon}
\label{sect:gevol} 

The first studies of the morphology-density relation of 
galaxies already demonstrated that this relation extends over a very wide range of densities
from high-density cores of rich clusters to isolated galaxies in low-global-density environments \citep{1974Natur.252..111E,
1980ApJ...236..351D, 1984ApJ...281...95P,
1987MNRAS.226..543E}. 
More recently, this was shown, for example, by \citet{2007ApJ...658..898P, 2012A&A...545A.104L, 
2014A&A...562A..87E, 2015MNRAS.451.3249A, 2016MNRAS.457.2287A}.

We find that the star-formation properties of galaxies depend both on their
local (group versus single galaxies, and rich groups versus poor groups) 
and global (high versus low) density environment. 
Single galaxies in our sample may be the brightest 
galaxies of faint groups in which other members of the groups
are outside of SDSS spectroscopic sample limits \citep{2009A&A...495...37T}.
In such poor, 
faint groups, or for single galaxies,
galaxy transformations are less effective than
in richer groups. This agrees with earlier results showing that the poorer
the group, the higher is the fraction of star forming galaxies
in it \citep{2014A&A...562A..87E}. 
This shows the importance of the group environment in which 
galaxies are transformed from BSF galaxies to VO galaxies.

The star-formation activity and other properties 
of BSF, RSF, and VO galaxies correlate quite well
with their local luminosity densities. 
In denser environment the current SF activity of galaxies is lower, galaxies are more concentrated and
have higher stellar velocity distributions. 
The properties of BSF, RSF, and VO galaxies (except the metallicities of VO galaxies,
see Table~\ref{tab:galpopc})
are statistically similar in both global environments.
The similarity of galaxy properties in the supercluster and in the cocoon 
is somewhat unexpected.
Figure~\ref{fig:dn}
suggests that transformations of galaxies are more efficient in the high global density
environment. Also, earlier studies have shown
that galaxies in poor groups in the low global density environment
need more time to evolve from star-forming galaxies to
quenched galaxies \citep{2012A&A...545A.104L}.
Owing to this, we could expect that within a given galaxy population
galaxies in high and low global density
environments could be somewhat different, but we find this for RQ galaxies only.

In our sample, BSF galaxies 
lie in the lowest local density $D1$ environments both in the supercluster
and in the cocoon. They also have the lowest stellar masses, as they are,
according to their properties, young late-type galaxies that are 
still growing their stellar mass.
In the cocoon, BSF galaxies can be found in outer parts of filaments. In the supercluster,
they lie mainly in the outer parts of the supercluster
main body. In both global environments. these galaxies may enter 
galaxy systems for the 
first time, as described in \citet{2019OJAp....2E...7A}. 
This may be the reason why they are similar in both environments.

In the
cocoon, in the similar, very low   local density environment, 
some late-type galaxies (BSF)
are in active star formation stage, but other late-type galaxies are in
already quenched stage (RQ).
Both BSF and RQ galaxies in the cocoon lie in the periphery of structures
(Fig.~\ref{fig:radecpop}).
Figure~\ref{fig:rqvorssm} shows that the stellar
masses for BSF galaxies with $log M^{\mathrm{*}} = 10.2$
are lower than those of RQ galaxies, which have $log M^{\mathrm{*}} = 10.6$.
According to models by \citet{2020ApJ...889..156C}, environmental quenching
efficiencies  for these masses are $0.5$ and $0.6$, respectively. This means that
the mass quenching efficiency is similarly  low for both. 
The difference in the star-formation properties of these galaxies 
can only be partly attributed to their different masses and there should be
also  other factors which
determine the star formation stage of galaxies. 
One possibility is that star formation in the present-day RQ galaxies started early
and they have had more time than BSF galaxies
to form their stars and increase their stellar mass.

Interestingly, we found RQ galaxies in the cocoon, but not everywhere in the low local density
outskirts of filaments and groups; that is, only in regions
with a declination below 22 degrees. The  number of RQ galaxies is small and
this may simply be due to this. 
It is also possible that the properties
of filaments, including their gas content, which could fuel the star formation in galaxies, 
vary from filament to filament. 
Earlier studies found a variety of star forming properties of galaxies 
in the infall zones of rich 
galaxy clusters, where some infalling structures are populated mostly 
by passive, red galaxies, and others by blue, star-forming galaxies
\citep{2010A&A...522A..92E, 2012MNRAS.421.1949V, 2016MNRAS.461.1202J, 2017A&A...607A.131D,
2018A&A...610A..82E, 2020A&A...638A.126D}. 
The variety of galaxy populations near rich clusters may be attributed to the 
different history of these galaxies \citep{2013MNRAS.430.3017B}.
This needs further study, together with the analysis 
of the galaxy and gas content of filaments 
from simulations, in order to clarify how large the variations of the properties
are for individual filaments. 

In the supercluster, RQ galaxies typically lie in the 
infall zones of substructures of the cluster and groups, in the merging zone of 
merging groups, and in the turnaround region of the supercluster main body 
\citep{2018A&A...610A..82E, 2018A&A...620A.149E}.  
The concentration indexes of RQ galaxies  in the supercluster are significantly lower
than in the cocoon.  
This hints that RQ galaxies in the cocoon and in the supercluster
may represent galaxies at different epochs of the evolution, and
processes responsible for star formation 
quenching in RQ galaxies in the cocoon may be somewhat different and more 
diverse from those in the supercluster. 
One possibility is that the quenching of star formation of 
RQ galaxies in the cocoon was due to their infall into primordial groups and detachment
of small scale primordial filaments feeding a galaxy with fresh gas 
\citep{2019OJAp....2E...7A}. 
In the supercluster, in the infall zones of groups and clusters
RQ galaxies may have several star formation events in the past,
which is analogous to what was suggested for galaxies with young stellar populations in the
nearby Universe by \citet{2020MNRAS.492.1791M}. Thus, they actually represent 
galaxy population, which is different from RQ galaxies in the cocoon.
However, this explanation ignores the finding that the stellar
masses of RQ galaxies are statistically similar in the supercluster and in
the cocoon.

RSF galaxies represent galaxies in transformation between BSF and 
VO galaxies. 
The   local densities 
$D1$ around RSF galaxies are between BSF and VO galaxies both in the supercluster
and in the cocoon.  
The distribution of RSF galaxies follows the distribution of groups 
and filaments, but they seem to avoid central parts of these systems
\citep[for RSF galaxies in the supercluster, this was shown in][]{2018A&A...620A.149E}. 
RSF galaxies are typically of late type
with stellar masses clearly higher than stellar masses of BSF galaxies, being 
comparable to those of RQ and VO galaxies. Their origin may be related to
an accretion into a filament or into a group or cluster without experiencing major
merger. This process is described as a cosmic web detachment without morphological 
change in \citet{2019OJAp....2E...7A}.
We may suppose that the timescales of star formation
quenching and the change in  morphology are different 
\citep{2019MNRAS.486..868K, 2019MNRAS.488.4117H}. This is in agreement
with studies which showed that the changes in colours of galaxies
and  in their morphology have different timescales
\citep{2010MNRAS.405..783M, 2011ApJ...736...51E}
A  number of physical processes have been related to galaxy transformations
in groups (the so-called preprocessing),
like starvation, overconsumption and others 
\citep[see, e.q. ][and references therein]{2009MNRAS.400..937M, 2014MNRAS.442L.105M,
2019OJAp....2E...7A, 2019MNRAS.490L...6D}.

\citet{2016MNRAS.457.2287A} found that 
among spiral galaxies in GAMA  fields
galaxies which lie closer to the filament axes have higher stellar masses 
and higher star-formation rates than 
spiral galaxies in the outer parts of filaments. 
These results partly agree with our findings about BSF and RSF galaxies,
and disagree with our results about RQ galaxies.
One possibility that would lessen this disagreement is that 
the time to be a RQ galaxy is very short
and we happen to evidence a snapshot of their life in some filaments only
\citep[see also][for discussion and references]{2017A&A...607A.131D}. 

We found also that in the cocoon there are surprisingly
many VO galaxies, even among single galaxies. 
VO galaxies lie in a wide range of local and 
global environments, with $D1$ from $ < 1$ in the cocoon to $> 800$ in the supercluster 
and $D8$ values also from $ < 1$ (cocoon) to $ > 20$ (supercluster).
The richness of their host groups vary from single galaxies to the richest 
cluster in our study, A2142, with  several orders of host group mass range,
$10^{13}h^{-1}M_\odot - 10^{15}h^{-1}M_\odot$ (see Table~\ref{tab:gr10}). 
Their properties are nearly independent of the global environmental 
density and, thus, the overall conditions for these galaxies to form and evolve 
had to be similar 
(or leading to similar galaxies) both in high- and low-global-density environments. 
This seems to be somewhat unexpected owing to the very different conditions
in rich and  poor groups.

For rich systems, both observations and simulations show that galaxy 
quenching starts in protogroups and protoclusters
at early epochs of group formation 
\citep{2019ApJ...887..183Z, 2019MNRAS.484.1702P}.
When a galaxy falls into a cluster, 
during the first $1.5-2.5$ Gyrs, 
star-formation quenching is slow. Thereafter, 
when a galaxy reaches to dense regions of the cluster, quenching 
proceeds more rapidly 
\citep{2019A&A...621A.131M, 2019ApJ...873...42R, 2020A&A...638A.133L, 2020ApJS..247...45R}.
Galaxies 
in poor groups in the low-global-density environment need more time to evolve from 
star-forming galaxies into quenched galaxies \citep{2012A&A...545A.104L}.
Single galaxies in both environments are least affected by the 
environment, and their evolution should be the slowest.
This means that large number of old and quenched galaxies in poor groups in the low-global-density
environment may be at odds with theoretical predictions \citep{2019MNRAS.490.3309M}.

\subsection{Connectivity of the supercluster and the evolution of SCl~A2142 in the cosmic web }
\label{sect:con}

\begin{figure}[ht]
\centering
\resizebox{0.44\textwidth}{!}{\includegraphics[angle=0]{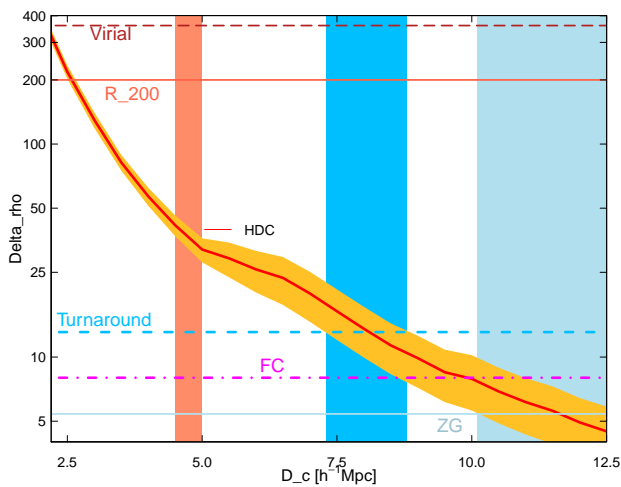}}
\caption{
Density contrast $\Delta\rho = \rho/\rho_{\mathrm{m}}$ versus
clustercentric distance $D_c$ for the SCl~A2142 main body (red line). 
Golden area shows error corridor
from mass errors. Characteristic density contrasts for the standard cosmological model
are
denoted as follows: 
$\Delta\rho = 360$ (virial),
$\Delta\rho = 200$ ($r_{200}$), $\Delta\rho = 13.1$ (turnaround,
blue dashed line),
$\Delta\rho = 8.73$ (future collapse FC, magenta dash-dotted line), 
and $\Delta\rho = 5.41$
(zero gravity ZG, light blue solid line).
Tomato, blue, and light  blue vertical areas mark borders
of the HDC of the supercluster, % (with $\Delta\rho \approx 30$), 
turnaround region of the supercluster 
main body, and zero gravity region. 
}
\label{fig:dcrho}
\end{figure}

\begin{figure}[ht]
\centering
\resizebox{0.47\textwidth}{!}{\includegraphics[angle=0]{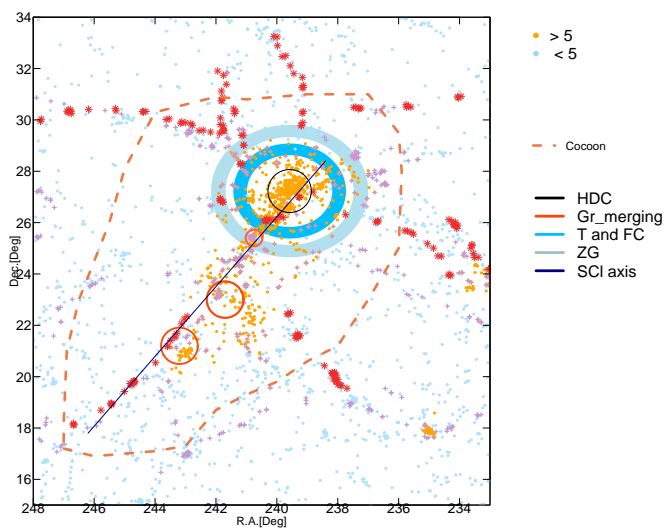}}
\caption{
Distribution of galaxies and filaments in the sky plane in and around the supercluster SCl~A2142. 
Colours denote galaxies in regions
of different luminosity-density as in Fig.~\ref{fig:radecvel}. 
Galaxies in short filaments are denoted with violet
colour, and galaxies in long filaments with length $ \geq 20$~\Mpc\
are denoted with dark red colour. 
HDC of the supercluster is marked with black circle.
Orange circles mark the location of merging 
groups which will separate from the supercluster in the future.
Blue stripe marks turnaround (T) region, 
and light blue circle shows borders of zero gravity (ZG)
region where long filaments are detached from the
supercluster (see text). Future collapse (FC) region 
lies between these regions.
Navy line denotes the supercluster axis.
Orange dashed line shows the cocoon boundaries.
}
\label{fig:radecevol}
\end{figure}

We begin the discussion about the connectivity and
possible evolution of SCl~A2142 in the cosmic web
with a short analysis of the possible evolution and dynamical state of the supercluster 
according to the spherical collapse model. The spherical collapse model describes the evolution 
and dynamical state of a 
spherical perturbation in an expanding universe.
The details of the model
with references are given in Appendix~\ref{sect:sph}.

In the spherical collapse model, the evolution of a spherical perturbation is determined by its 
density contrast (Appendix~\ref{sect:sph}).
In the process of evolution. perturbations have several important epochs.
For each of them, there is a corresponding characteristic density 
contrast. The first epoch  is called the 'turnaround', with
a density contrast of $\Delta\rho_{T} = 13.1$.
Turnaround is defined as the epoch 
at which a spherical overdensity region
decouples from expansion and its collapse begins.
Overdensity regions in which density contrast is not sufficient
to   collapse at present may eventually
experience turnaround and collapse in the future, if their current overdensity is
$\Delta\rho_{FC} = 8.73$
(where FC stands for future collapse).
The density contrast $\Delta\rho_{ZG} = 5.41$ corresponds to a so-called zero 
gravity 
radius (ZG) where  gravitation equals  expansion and
 which borders the region which may stay gravitationally bound
\citep{2015A&A...577A.144T, 2015A&A...581A.135G}.

In Fig.~\ref{fig:dcrho}, we plot the
density contrast $\Delta\rho$ with respect to the distance $D_c$
(the distance from the supercluster centre at the cluster A2142)
for the supercluster SCl~A2142.
Density contrast $\Delta\rho$ is calculated using mass estimates 
as explained in Appendix~\ref{sect:sph}.
We mark in Fig.~\ref{fig:dcrho} characteristic density 
contrasts (and their characteristic radii) listed above, 
namely.
the turnaround region of the supercluster, the future collapse  
region (FC), and the zero gravity (ZG) region. We additionally indicate the border 
of the HDC of the supercluster with density contrast $\Delta\rho \approx 30$. 
We show these regions also in Fig.~\ref{fig:radecevol} 
where we plot the sky distribution of galaxies and filaments
in SCl~A2142 and in its cocoon.

To start the analysis of a possible dynamical state and evolution of
the supercluster from inside out, the first region highlighted in Figs.~\ref{fig:dcrho}
and \ref{fig:radecevol} is the HDC of the supercluster with the 
density contrast
at its borders $\Delta\rho \approx 30$, and radius of $R_{HDC} \approx 5$~\Mpc.
The high density contrast suggest   
that the HDC has passed turnaround with the density contrast 
$\Delta\rho = 13.1$, and continues contracting. 
This is supported by the analysis of the structure and dynamical state
of the HDC in Sect.~\ref{sect:fil}  which suggested
that groups in the HDC are falling into the cluster A2142.

The connectivity of the cluster A2142 $\pazocal{C} = 6 - 7$.
The connectivity 
within HDC may have changed during the evolution.
For example, groups Gr5 and M1 (see Fig.~\ref{fig:hdcradec} and Table~\ref{tab:gr10} 
may be remnants
of a small filament near the cluster A2142 destroyed by the infall
\citep{2018A&A...620A.149E}. 
They may be  related to the radio ridge detected by \citet{2017A&A...603A.125V}.
Short filaments and substructures which surround the cluster A2142
in the HDC are infalling into the cluster. They are detached from the main body of the supercluster
by a minimum in the galaxy distribution
\citep[this is seen also in Fig.~5 in][as a minimum in the density distribution]
{2015A&A...580A..69E}. We suggest that this is a signature of
a collapse of the HDC. \citet{2018A&A...610A..82E}
proposed that the collapse may have started at least $4$~Gyrs ago, 
approximately at redshift $z \approx 0.5$. 
Simulations show that this redshift correspond to half-mass period in the
evolution of rich clusters \citep{2015JKAS...48..213K}.

Figures~\ref{fig:dcrho} and \ref{fig:radecevol} show that the radius of the turnaround region
of the supercluster (shown in dark blue), $R_T \approx 7 - 9$~\Mpc. 
This region is populated mostly by poor groups and single
galaxies in short filaments. In Sect.~\ref{sect:con}, we showed that at 
clustercentric distances of $D_c \geq 7,$ six
long filaments begin, which extend out of the supercluster.
This is the same distance at which
galaxies with young stellar populations appear in long filaments
(Sect.~\ref{sect:galprop}).
At these clustercentric distances,
\citet{2018A&A...620A.149E} detected an excess of star-forming and recently
quenched galaxies. 
We propose that long filaments do not reach 
the inner regions of the supercluster because of the 
collapse of the supercluster main body, which have destroyed the structures in this 
region and may affect the star--formation properties  of galaxies.

According to the predictions of the spherical collapse model. the
outer parts of the supercluster main body with radius 
$R_{FC} \approx 9$~\Mpc\ will collapse in the future. 
Zero gravity region with a radius of 
$R_{ZG} \approx 10 - 13$~\Mpc\ approximately surrounds the supercluster
main body. 
This means that regions at larger distances from the supercluster centre
will not become gravitationally bound. This agrees with an earlier
analysis which showed that groups in the supercluster tail will
probably separate from the supercluster in the future
\citep{2015A&A...580A..69E, 2018A&A...620A.149E}.
The radius of a region which corresponds to the 
linear mass scale is $R_{L} \approx 15$~\Mpc\
\citep[see][]{2015A&A...581A.135G}.
This region
approximately follows the cocoon borders where the densities are the lowest.
We do not have galaxy velocity data for the SCl~A2142 region but simulations
show that the lowest densities correspond to the lowest values of the peculiar 
velocities of galaxies.

We emphasise that the characteristic  radii 
in Fig.~\ref{fig:dcrho} were found using mass
estimates  of groups as described in Appendix~\ref{sect:sph}. 
These radii  are in a good accordance with those
seen from filament distribution. 
This coincidence supports our interpretation of the disconnection of long filaments.
Also, as noted in \citet{2015A&A...580A..69E},
the mass estimates of SCl~A2142 agree well with those
found for supercluster masses from simulations 
\citep{2014A&A...567A.144C} which means that our estimates of characteristic 
radii are not strongly affected by mass errors.

As discussed in Sect.~\ref{sect:conn}, 
the results about the connectivity depend on how filaments
are defined, and on the parameters chosen for filament membership
(in our case, the distance of a galaxy to 
the nearest filament axis, $D_{fil}$). 
Figure~\ref{fig:dfil} shows that in our study the choice of $D_{fil}$
value for filament membership, $D_{fil} \leq 0.5$~\Mpc\ is justified.
This choice enables to study fine details in the distribution
of groups and filaments inside the supercluster, and especially in its HDC.
The situation may be different in studies which focus on the overall
properties of the cosmic web, where detection of wider
filaments may be preferred \citep[see, for example, ][]{2020arXiv200201486M}.

We also found that galaxy content along the longest filaments does not change 
significantly with the distance from the supercluster centre.
This 
is somewhat incompatible with the theoretical calculations 
by \citet{2013MNRAS.430.3017B} who showed that  galaxy infall 
into a cluster occurs along  filaments and there should be a significant 
dependence of the fraction of gas rich galaxies with a distance from the cluster. 
The fraction should decrease $3-4$ times when the distance 
changes from $(4 \rightarrow 1.5)\, R_{200}$ or from $11$~\Mpc\ to $5$~\Mpc. 
This also disagrees with the recent observational findings about galaxy properties in
filaments \citep{2020MNRAS.491.4294K}.
The reason of this discrepancy may be related to the dynamical evolution 
of the supercluster, and the detachment of filaments.
Inside the HDC, red, passive galaxies are closer to the cluster A2142 than blue,
star-forming galaxies \citep{2018A&A...620A.149E}, 
in agreement with 
\citet{2008MNRAS.388.1152P, 2013MNRAS.430.3017B,
2020MNRAS.491.4294K}.
At the borders of the collapsing region of the supercluster, 
long filaments are detached, and this also affects galaxy properties.
There is an excess of star-forming at the
turnaround region. Outside of the supercluster, the dependence 
of galaxy properties along filaments with regard to the distance from the supercluster is weak.

The minimum in the galaxy distribution
around the HDC is similar to the minimum which have been noted around some other
rich galaxy clusters, for example, around the cluster A1436 in the Ursa Major supercluster
\citep{2007AstL...33..211K, 2012A&A...542A..36E}. It is 
possible that galaxies near cluster have already merged with the cluster.
Interestingly, we found similar 'distance gap' around one poor galaxy group
in the cocoon. We denote this as group $G_{\mathrm{gap}}$ (Gr7481 in the catalogue).
$G_{\mathrm{gap}}$ has also the largest magnitude gap between its two brightest
galaxies among cocoon groups, $|\Delta M_{12}| = 1.45$. 
This group lies near the endpoint of filament. 
We show the distribution of galaxies in this group and
its PPS diagram in Fig.~\ref{fig:radecvelpps7481}.

Figure~\ref{fig:radecvelpps7481} shows that galaxies in the centre of 
the group are all with old stellar populations.
Star-forming galaxies, including one RSF galaxy, lie in the outskirts of
the group. This may be a signature of enhanced star formation
due to the infall into the group. 
The nearest galaxies to the group are also RSF galaxies.
There  are no 
galaxies in the close environment of this group in the interval
of groupcentric distances of approximately $1 - 4$~\Mpc. 
It may be possible that galaxies formerly near $G_{\mathrm{gap}}$ have
already become the members of the group, and the accretion occurred along the filament
where $G_{\mathrm{gap}}$ lies. The brightest galaxy in this group 
may have merged some group members, and this may be the origin of its
large magnitude gap. We note that for this group, $|\Delta M_{13}| = 2.25,$ 
and it might be considered as fossil group candidate.
However, there are no available X-ray observations in
and around this group to confirm it as an X-ray source.
Also, we do not expect to see the X-ray
emission of fossil groups at z=0.1 in RASS.
These examples show one epoch in the evolution of galaxy groups and clusters 
and their connectivity, and deserves a special study which should include a larger sample
of groups and clusters.

\begin{figure}[ht]
\centering
\resizebox{0.23\textwidth}{!}{\includegraphics[angle=0]{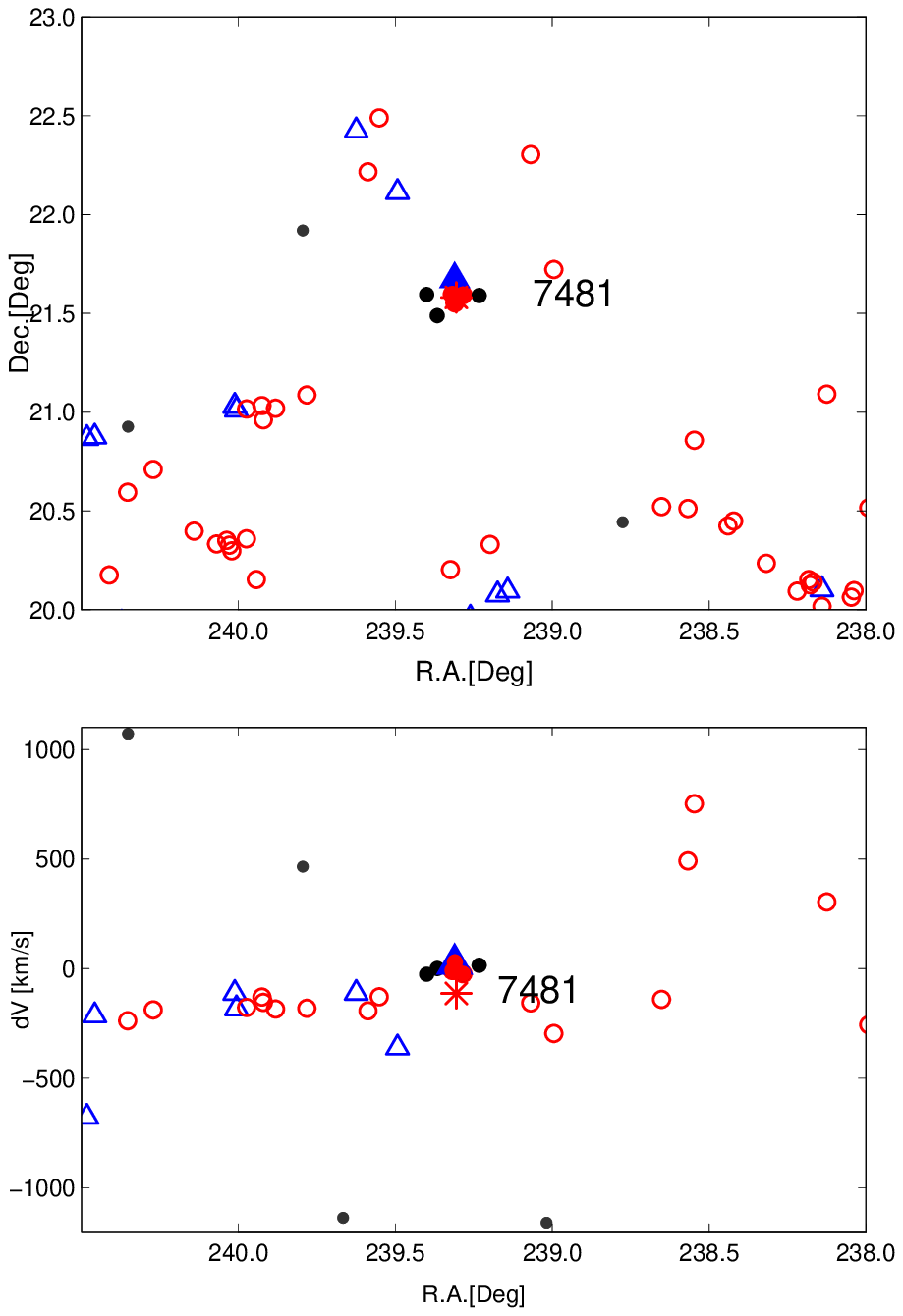}}
\resizebox{0.23\textwidth}{!}{\includegraphics[angle=0]{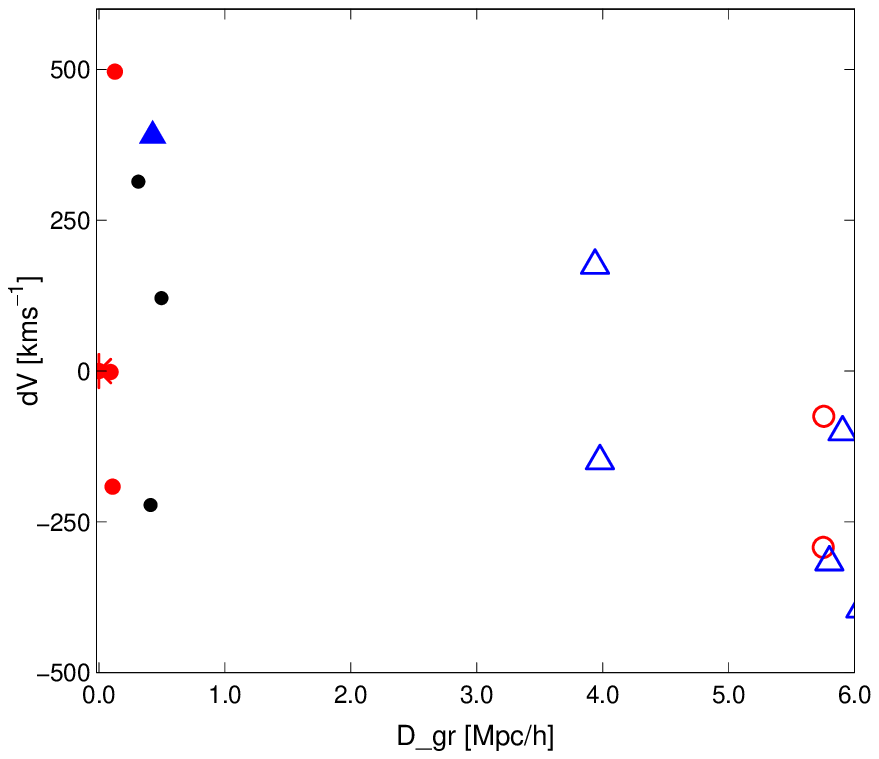}}
\caption{
Sky distribution, and R.A. - velocity difference distribution for galaxies
centred at the group $G_{\mathrm{gap}}$ (left panels),
and PPS diagram for $G_{\mathrm{gap}}$ and its neighbourhood galaxies (right panel). 
Red circles denote VO galaxies, and blue triangles denote
RSF galaxies. 
Filled symbols show group members, and empty symbols mark galaxies
in the group neighbourhood.
Other galaxies (blue star forming galaxies and red galaxies with stellar populations
of intermediate age) are marked with grey dots.
}
\label{fig:radecvelpps7481}
\end{figure}

The presence of merging groups in the supercluster
tail is another
evidence of dynamical activity in the supercluster
\citep{2018A&A...620A.149E}. It is also likely that merging groups in the supercluster
tail will separate from the supercluster during future evolution
\citep{2015A&A...580A..69E, 2015A&A...581A.135G}.
This suggests that the growth of the whole supercluster has stopped
and the split to several systems has started. Therefore,
the borders of the SCl~A2142 cocoon also will change in the future
and there will be at least three cocoons where now there is one.

Our study shows that the supercluster SCl~A2142 together with 
galaxy filament at the extension of the supercluster forms the longest
straight structure found from observations so far. 
Such systems can be used as a cosmological probe
to test whether 
such straight structures in the cosmic web are consistent with the $\Lambda CDM$
model.  This problem has not yet been studied implicitly.
\citet{2012ApJ...759L...7P} analysed the sizes of the 
most massive structures identified in the mock catalogues based on
Horizon II simulations of a structures formation in the $\Lambda CDM$
Universe. This study showed that the largest systems from simulations
have similar sizes as the largest systems of galaxy clusters from 
SDSS DR7 catalogue. Interestingly,
in Fig. 2 by \citet{2012ApJ...759L...7P}, one of the largest systems 
from mock catalogues
is visually very similar to SCl~A2142, having spherical body with long tail, 
although this system is three times larger than SCl~A2142. 
This cannot be taken as direct comparison between SCl~A2142
and mock catalogues since the systems were chosen differently.
However, it might be possible that straight systems can be found in very
large simulations. This shows that when comparing observations and simulations,
morphological and topological measures
of the cosmic web elements should be used in addition to other statistics
\citep[][and references therein]{2011A&A...532A...5E, 2020MNRAS.493.5972H}.

\section{Summary}
\label{sect:sum} 

We determined the boundaries of the supercluster SCl~A2142 cocoon using the luminosity-density 
field and studied the galaxy, group, and filament content therein.
Our study showed that the connectivity of the supercluster
and the galaxy content of groups and filaments are related to the 
evolution of the supercluster and its cocoon. Our results
can be summarised as follows.

\begin{itemize}
\item[1)]
The supercluster A2142, together with the long filament connected to it, form
the longest straight structure in the Universe detected so far, with a 
length of approximately $75$~\Mpc.
\item[2)]
The radius of SCl~A2142 cocoon, determined as the minimum in the luminosity-density field around the supercluster, is approximately equal to the
linear expansion scale, $\approx 20$~\Mpc. The cocoon is elongated, with the length
of approximately $80$~\Mpc.
\item[3)]
The connectivity of the cluster A2142 is $\pazocal{C} = 6 - 7$. 
The supercluster main body with six long surrounding filaments
has $\pazocal{C} = 6$. The connectivity of poor groups in 
the cocoon and supercluster is $\pazocal{C} = 1 - 2$.
These numbers are compatible with theoretical calculations by \citet{2018MNRAS.479..973C} 
for the standard $\Lambda$CDM model for corresponding group and cluster masses. 
\item[4)]
The supercluster  main body
is collapsing and the long filaments 
are detached from it  at the turnaround region of the supercluster main body.
These results suggest that the supercluster may split into
several systems in the future. Also, its cocoon will change in their future evolution.
\item[5)]
We found various trends with the 
local (group and cluster) and global (supercluster and cocoon) environment
in the properties of galaxies and groups.
\subitem{5.1)}
While groups in a supercluster contain higher percentage of red galaxies and galaxies with
high values of $D_n(4000)$ than in the cocoon, 
the star-formation properties of single galaxies 
are similar in all environments.
\subitem{5.2)}
Local luminosity-density at the location of poor groups in the cocoon is lower than
that at poor groups in the supercluster. 
Local luminosity-density is the lowest around blue star-forming galaxies, and the 
highest around galaxies with old stellar populations (those with $D_n(4000) \geq 1.75$).
\subitem{5.3)}
Galaxies with the oldest stellar populations 
occupy groups with masses in a wide interval of 
$10^{13}h^{-1}M_\odot - 10^{15}h^{-1}M_\odot$.
They have local and global densities in a range
from $D1$ and $D8 < 1$  in the cocoon  
to $D1 > 800$ ($D8 > 20$) in the supercluster. 
There are VO galaxies also among single galaxies in both environments.
\subitem{5.4)}
Recently quenched galaxies lie in the low-density outskirts of some galaxy systems,
but not everywhere, and their properties in the cocoon and in the supercluster 
are different.
\subitem{5.5)}
There is no systematic change in galaxy properties along
long filaments with the distance from the supercluster centre,
which cannot be explained with the changes in local densities (presence of groups)
or global densities.
\end{itemize}

The study of 
the supercluster SCl~A2142, galaxies and galaxy groups in it
and in its environment,
together with the earlier study of SCl~A2142  \citep{2018A&A...620A.149E},
shows how the supercluster and structures around it evolve. 
We detected large differences between the galaxy content of individual groups and filaments.
Together with earlier findings about large variations in the galaxy
content of infalling structures around galaxy clusters this deserves
future analysis.
Our study suggests  that the reasons why galaxies are recently quenched are
different in the supercluster and in the cocoon.
Unexpectedly, VO galaxies, which lie in a wide range of systems from single galaxies 
in the cocoon to the richest cluster of the supercluster, have a similar concentration index,
stellar velocity dispersion, and morphology
in all these rather extreme environments. 
Since local conditions, which
affect galaxies in the centre of rich cluster, are very different from those
in poor groups or for single galaxies, we might expect differences in their properties. 
Further studies are needed to clarify the details of galaxy evolution in various
environments.

The supercluster has 
a long, straight filament as the extension of the supercluster,
forming the longest straight structure detected so far.
The presence of such structures may be a challenge for the theories
of the structure formation in the Universe. We plan to address this question 
in a future studies.

\begin{acknowledgements}
We thank the referee for a very detailed report which helped us to improve the paper.
We thank Enn Saar, Dmitri Pogosyan, Miguel Aragon-Calvo, Rien van de Weygaert,
Rain Kipper, and Teet Kuutma for useful discussions.
We are pleased to thank the SDSS Team for the publicly available data
releases.  Funding for the Sloan Digital Sky Survey (SDSS) and SDSS-II has been
provided by the Alfred P. Sloan Foundation, the Participating Institutions,
the National Science Foundation, the U.S.  Department of Energy, the
National Aeronautics and Space Administration, the Japanese Monbukagakusho,
and the Max Planck Society, and the Higher Education Funding Council for
England.  The SDSS website is \texttt{http://www.sdss.org/}.
The SDSS is managed by the Astrophysical Research Consortium (ARC) for the
Participating Institutions.  The Participating Institutions are the American
Museum of Natural History, Astrophysical Institute Potsdam, University of
Basel, University of Cambridge, Case Western Reserve University, The
University of Chicago, Drexel University, Fermilab, the Institute for
Advanced Study, the Japan Participation Group, The Johns Hopkins University,
the Joint Institute for Nuclear Astrophysics, the Kavli Institute for
Particle Astrophysics and Cosmology, the Korean Scientist Group, the Chinese
Academy of Sciences (LAMOST), Los Alamos National Laboratory, the
Max-Planck-Institute for Astronomy (MPIA), the Max-Planck-Institute for
Astrophysics (MPA), New Mexico State University, Ohio State University,
University of Pittsburgh, University of Portsmouth, Princeton University,
the United States Naval Observatory, and the University of Washington.

The present study was supported by the ETAG projects 
IUT26-2, IUT40-2,  PUT1627, by the European Structural Funds
grant for the Centre of Excellence "Dark Matter in (Astro)particle Physics and
Cosmology" (TK133), and by MOBTP86.
GC acknowledges support by the Deutsches Zentrum für Luft-
und Raumfahrt under grant No. 50 OR 1905, and
BD acknowledges the support of the Czech Science Foundation grant
19-18647S and the institutional project RVO 67985815.
This work has also been supported by
ICRAnet through a professorship for Jaan Einasto.

We applied in this study, the R statistical environment 
\citep{ig96}.

\end{acknowledgements}

\bibliographystyle{aa}
\bibliography{scl2142env.bib}

\begin{appendix}

\section{Spherical collapse model and characteristic density contrasts}
\label{sect:sph} 

The standard cosmological model 
states that the formation and evolution of the structure in the Universe is governed by the
gravitational attraction of dark matter and the repulsion of dark energy.
The evolution and dynamical state of a 
spherically symmetric perturbation 
can be described with a spherical collapse model.
We use this model to find characteristic scales for SCl~A2142 that are related to its
evolution. 
According to this model,
the dynamical state of a perturbation is determined by the mass in its interior
\citep{1972ApJ...176....1G, 1980lssu.book.....P},
usually described using the density contrast.
The density contrast of spherical perturbation is defined as
\begin{equation}
\Delta\rho = \rho/\rho_{\mathrm{m}},
\end{equation} 
where 
$\rho = M/V$ denotes the matter density in the volume $V$ with mass $M$.
$\rho_{\mathrm{m}} = \Omega_{\mathrm{m}}\rho_{\mathrm{crit}} =
 3 \Omega_{\mathrm{m}} H_0^2 / 8\pi G$ 
is the mean matter density in the local universe
\citep[see ][for details and references]{2015A&A...581A.135G, 2015A&A...575L..14C}.
Density contrast $\Delta\rho$ for a spherical volume $V = 4\pi R^3/3 $ 
is
\begin{equation}
\Delta\rho = 0.86\times 10^{-12}\Omega_{\mathrm{m}}^{-1}~(\frac{M}{h^{-1}M_\odot})~(\frac{R}{h^{-1}~{\mathrm{Mpc}}})^{-3}.
\label{eq:sphrho}
\end{equation} 

The mass of a structure inside a sphere can be found using
Eq.~(\ref{eq:sphrho}):
\begin{equation}
M(R) = 1.16\times 10^{12}~\Omega_{\mathrm{m}}\Delta\rho~
(R/h^{-1}~{\mathrm{Mpc}})^{3}h^{-1}M_\odot. 
\label{eq:mass1}
\end{equation}

The spherical collapse model defines several epochs in
the evolution of a perturbation, each of which has a characteristic density contrast. 
Corresponding density contrasts 
are marked in Fig.~\ref{fig:dcrho}.
The turnaround (T) corresponds to the epoch 
at which a spherical overdensity region
reaches its maximum expansion and the collapse begins. 
The density contrast at turnaround $\Delta\rho_{T} = 13.1$ \citep[for the cosmological
parameters used in this paper, see, for example, ][]{2015A&A...581A.135G}. 
In cosmology without dark energy the density contrast
at turnaround $\Delta\rho_{T} = 5.5$ \citep{2002sgd..book.....M}.
Overdensity regions which 
turnaround and collapse in the future (FC) have present overdensity 
$\Delta\rho_{FC} = 8.73$ \citep{2015A&A...575L..14C}.
The density contrast $\Delta\rho_{ZG} = 5.41$ corresponds to so-called zero 
gravity (ZG), 
at which the radial peculiar velocity component of 
test particle velocity equals the Hubble expansion
and the gravitational attraction of the system and its expansion are equal.
The density contrast 
$\Delta\rho = \rho/\rho_{\mathrm{m}} = 1$ 
corresponds to linear mass scale or the Einstein-Straus radius at which
the radial velocity around a system reaches the Hubble velocity,
$u = HR$ and peculiar velocities $v_{\mathrm{pec}} = 0$
\citep{2015A&A...577A.144T, 2015A&A...581A.135G}. 
This scale approximately corresponds to the cocoon boundaries.

Additionally, we show in Fig.~\ref{fig:dcrho} the density contrast
$\Delta\rho_{200} = 200$. In galaxy group and cluster studies
the density contrast $\Delta\rho_{200}$ is often used to determine
group or cluster radius, called as $R_{200}$. 
In cosmology without dark energy, the density contrast
which corresponds to the virial radius of the cluster
has a value close to this, $\Delta\rho_{vir} = 178$
\citep{2002sgd..book.....M, 2005RvMP...77..207V},
However, often $R_{200}$
is used without specifying how it is defined
\citep[see short discussion in][]{2005RvMP...77..207V}. 
In the standard $\Lambda$CDM model, the virial density contrast 
is $\Delta\rho_{vir} = 360$ \citep{2015A&A...581A.135G}.
This is also marked in Fig.~\ref{fig:dcrho}.
In units of critical density, the density contrast at virial radius 
of a cluster is
$\Delta\rho_{vir} = \rho/\rho_{\mathrm{crit}} = 97$ \citep{2015A&A...581A.135G}.

To calculate the mass and to estimate mass errors
for the supercluster 
in Fig.~\ref{fig:dcrho}, 
we followed the procedure from \citet{2018A&A...620A.149E}.
In this study the dynamical mass of groups with at least four member galaxies was taken 
from \citet{2014A&A...566A...1T}. Mass estimates of very poor groups with 
less than four galaxies have very large scatter. For this reason we used 
their median mass. To obtain the total mass of very poor groups, this median
mass was multiplied by the number of groups.
To calculate errors, 50\% mass errors were
used.

\end{appendix}

\end{document}